\documentclass[prd,twocolumn,showpacs,groupedaddress,superscriptaddress,amsmath,amssymb]{revtex4-2} 
 
\newcommand{\vect}[1]{\boldsymbol{#1}}

\usepackage[utf8]{inputenc}
\usepackage[english]{babel}										
\usepackage{cmap}
\usepackage{bbm}
\usepackage[T2A]{fontenc}
\usepackage{enumerate}
\usepackage{multirow}
\usepackage{float}

\usepackage{amsmath,amsfonts,amssymb,mathtools}	
\usepackage{euscript} 
\usepackage{mathrsfs}
\usepackage{mathtext}

\usepackage{color}			
\usepackage{graphicx}
\usepackage[export]{adjustbox}
\DeclareGraphicsExtensions{.png,.jpg}
\graphicspath{}	
\usepackage[section]{placeins} 

\usepackage{physics}
\usepackage{braket}
\usepackage{xparse}	
\usepackage{slashed}

\def\nn{\nonumber}

\usepackage{indentfirst} 
\usepackage{enumitem}

\usepackage[colorlinks=true, filecolor=blue, citecolor=blue,urlcolor=black]{hyperref}
\usepackage{url}
\usepackage{cleveref}
\date{\today}

\begin{document}

\title{Probing hidden spin-2 mediator of dark matter with NA64$e$, LDMX, NA64$\mu$ and M$^3$} 

\author{I.~V.~Voronchikhin}
\thanks{Corresponding author}
\email[\textbf{e-mail}: ]{ivv1211@yandex.ru}
\affiliation{ Tomsk Polytechnic University, 634050 Tomsk, Russia}

\author{D.~V.~Kirpichnikov}
\email[\textbf{e-mail}: ]{dmbrick@gmail.com}
\affiliation{Institute for Nuclear Research, 117312 Moscow, Russia}

\begin{abstract}
The connection between Standard Model (SM) particles and dark matter (DM) can be introduced 
via hidden spin-2 massive mediator.  In the present paper we consider the simplified 
benchmark link between charged lepton sector of SM and DM particles which are 
considered to be a hidden Dirac fermions from the dark sector.  The regarding couplings  are 
established through the dimension-5 operators involving spin-2 mediator field and  the  
energy-momentum tensors of both SM and DM sectors.  We study in detail the implication of  this scenario for the 
lepton fixed-target  facilities, such as NA64$e$, LDMX, NA64$\mu$ and M$^3$. In particular, for the specific experiment 
we discuss in detail the missing-energy signatures of spin-2 boson production followed by its invisible decay into stable 
DM pairs. Moreover, we  derive the expected reaches of these experiments for the projected statistics of the
leptons accumulated on the target.     We also discuss the implication of both nuclear and atomic form-factor 
parametrizations for the differential spectra of hidden spin-$2$ boson emission, 
 the total cross-section of its production and
 the experimental reach of the fixed-fixed target facilities for probing hidden spin-$2$ DM mediator.
\end{abstract}

\maketitle


\section{Introduction and framework}

The nature of the dark matter (DM) particles remains puzzling for decades. The indirect evidences for the  DM 
are associated with galaxy rotation velocities, large scale structures, cosmic microwave background anisotropy, 
gravitational lensing, etc. Some extensions of the standard model (SM) imply connection between SM and DM via an idea of 
portals. For instance, the dark photon portal~\cite{Holdom:1985ag,Okun:1982xi,Boehm:2003hm,Pospelov:2007mp}
and Higgs boson portal~\cite{McDonald:1993ex,Burgess:2000yq,Wells:2008xg,Schabinger:2005ei}. 
Such portal scenarios suggest a systematic probing of DM and also provide a novel specific experimental signatures.
However, recently a scenarios of a massive spin-2 particle  as the mediator between DM and SM have been  
discussed~\cite{lee:2014GMDM,Kang:2020-LGMDM,Folgado:2019gie,Kang:2020yul,Lee:2014caa} in the context of gravity
model with warped extra-dimensions
~\cite{Han:2015cty,Dillon:2016fgw,Dillon:2016tqp,Carrillo-Monteverde:2018phy,Kraml:2017atm,Rueter:2017nbk,Folgado:2019sgz,Bernal:2018qlk}.
Furthermore, recently confirmed 4.2$\sigma$ discrepancy in the anomalous magnetic moment measurement of the muon \cite{PhysRevLett.126.141801} with respect to its theoretical prediction \cite{AOYAMA20201}:
\[
\Delta a_{\mu} = a_{\mu} (exp) - a_{\mu}(th) = (251 \pm 59) \times 10^{-11},
\]  can be explained by the one-loop effects induced by a hidden massive 
spin-2 particle~\cite{Kang:2020-LGMDM,Huang:2022zet}. Let us consider
now the couplings between the charged lepton sector of  the SM and the DM particles that
can be described by the benchmark  simplified  Lagrangian \cite{Kang:2020-LGMDM} involving 
dimension-5 operators of the energy momentum tensor and massive $G_{\mu \nu}$ field:
\begin{align}
 &  \mathcal{L} \supset       \frac{c_{\gamma}}{\Lambda} G^{\mu \nu} 
    \left(
        \frac{1}{4} \eta_{\mu \nu} F_{\lambda \rho} F^{\lambda \rho} 
    +   F_{\mu \lambda} F^{\lambda}_{ \nu}
    \right) \nn \\
& -
   \sum_l  \frac{i c_{l}}{2\Lambda} G^{\mu \nu}
    \left(
            \overline{l} \gamma_{\mu} D_{\nu} l 
        -   \eta_{\mu \nu} \overline{l} \gamma_{\rho} D^{\rho} l
    \right)
+ 
    \frac{c_{\chi}}{\Lambda} G^{\mu \nu} T^{\chi}_{\mu \nu},
\label{BechmarkLagrOftheModel1}        
\end{align}
where $F_{\mu \nu}$ is stress tensor for the photon field, $l$ is the label for
the charged leptons $(l=e,\mu)$,  $\Lambda$ is dimensional parameter for spin-2 interactions,  
$c_{\gamma}$, $c_{l}$, $c_{\chi}$ are the  dimensionless coupling constants for  the electromagnetic 
field, charged leptons of SM and DM sector  respectively; $T^{\chi}_{\mu \nu}$
is the energy momentum tensor of DM particles.  To be more concrete we address DM as  hidden Dirac fermion $\chi$ with 
mass of $m_\chi$.    In addition, it is worth noting that the coupling constants in the regarding scenario are assumed to 
be  independent. 

\begin{figure*}[!tbh]
\centering
\includegraphics[width=0.7\textwidth]{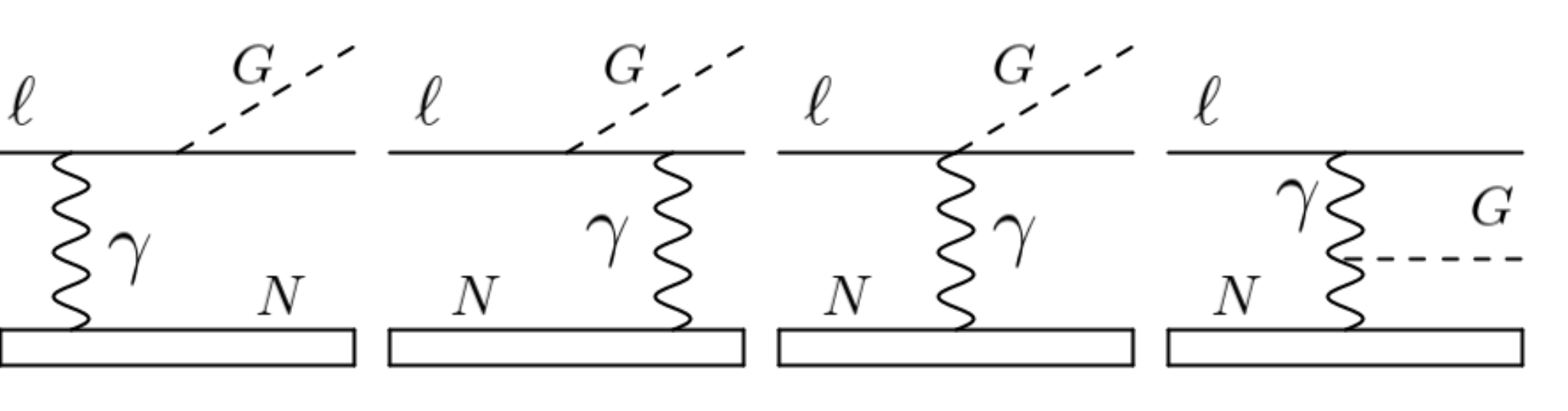}
\caption{Feynman diagrams describing missing energy signatures. 
\label{eNToeNGDiagram} }
\end{figure*}

In the present paper we show that corresponding setup~(\ref{BechmarkLagrOftheModel1}) has a very broad 
phenomenological implication and can be examined through the missing energy signatures in the existent 
(NA64$e$~\cite{Gninenko:2016kpg,NA64:2016oww,NA64:2017vtt,Gninenko:2018ter,Banerjee:2019pds,Dusaev:2020gxi,Andreev:2021fzd,Andreev:2022hxz,NA64:2022rme,Arefyeva:2022eba,Zhevlakov:2022vio,NA64:2021acr,NA64:2021xzo} and
NA64$\mu$~\cite{PhysRevD.105.052006,Kirpichnikov:2021jev}) and the projected 
(LDMX~\cite{PhysRevD.99.075001,Ankowski:2019mfd,Schuster:2021mlr,Akesson:2022vza} and
M$^3$~\cite{Capdevilla:2021kcf,Kahn:2018cqs}) 
lepton fixed-target experiments.   These signatures can be described by the bremsstrahlung-like reaction of the production
of a spin-2 boson by the charged lepton $l^-$ impinging  on a nucleus $N$. The corresponding diagrams are shown in 
Fig.~\ref{eNToeNGDiagram}. In the present paper we will focus on mainly invisible channel of dark spin-2 boson decay into DM 
particles, such that 
$\mbox{Br}(G\to \chi \bar{\chi})\simeq 1$. Furthermore, 
for the $G$-boson production  cross-section calculation   we exploit the equivalent photons approximation, also known as the
Weizsacker–Williams (WW) approach  that is typically exploited  
for the hidden particle yield estimate at both  beam dump and fixed-target experiments. In particular,
this approach provides a fairly reasonable approximation (i.~e.~at the level of $\lesssim 2\%$)
for the exact-tree-level production cross-sections of both hidden spin-$0$ and 
spin-$1$ bosons~\cite{PhysRevD.95.036010,PhysRevD.96.016004,Kirpichnikov:2021jev}.

In addition, we also discuss in detail the impact of both nuclear and atomic form-factor parametrizations on:  
(i) the differential spectra of $G$ boson emission, (ii) the total cross-section of its production 
(iii) the experimental reach of the fixed-fixed target facilities for probing hidden spin-2 boson.  

The paper is organized as follows. In Sec.~\ref{WW_approach_CS_Section} we derive explicitly  the double differential 
cross-section of $G$-boson production in WW approach. In Sec.~\ref{SectionFluxDetails} we discuss the set of 
form-factors that are typically used for the calculation of regarding cross-sections. In this section we 
also study  the impact of various form-factor parametrizations on the virtual photon flux distribution from the 
charged particles. In Sec.~\ref{ExperimentalBenchmark} we provide a description of the missing energy signatures 
for the analysis of DM production at lepton fixed-target experiments, such as NA64$e$, LDMX, NA64$\mu$ and~M$^3$.
In Sec.~\ref{CS_WW} we discuss the impact of form-factor parametrization on both differential and total cross-sections
of hidden spin-2 boson production. In Sec.~\ref{SectionExpectedReach} we obtain the  
constraints on the parameter space of spin-2 DM mediator from the NA64$e$, LDMX, NA64$\mu$ and~M$^3$ facilities.
In this section we also study implication of the form-factors for the expected reach of the regarding experiments.

\section{ The production cross section of DM mediator in WW approach
\label{WW_approach_CS_Section}
}

\begin{figure*}[!t]
\centering
\includegraphics[width=0.8\textwidth]{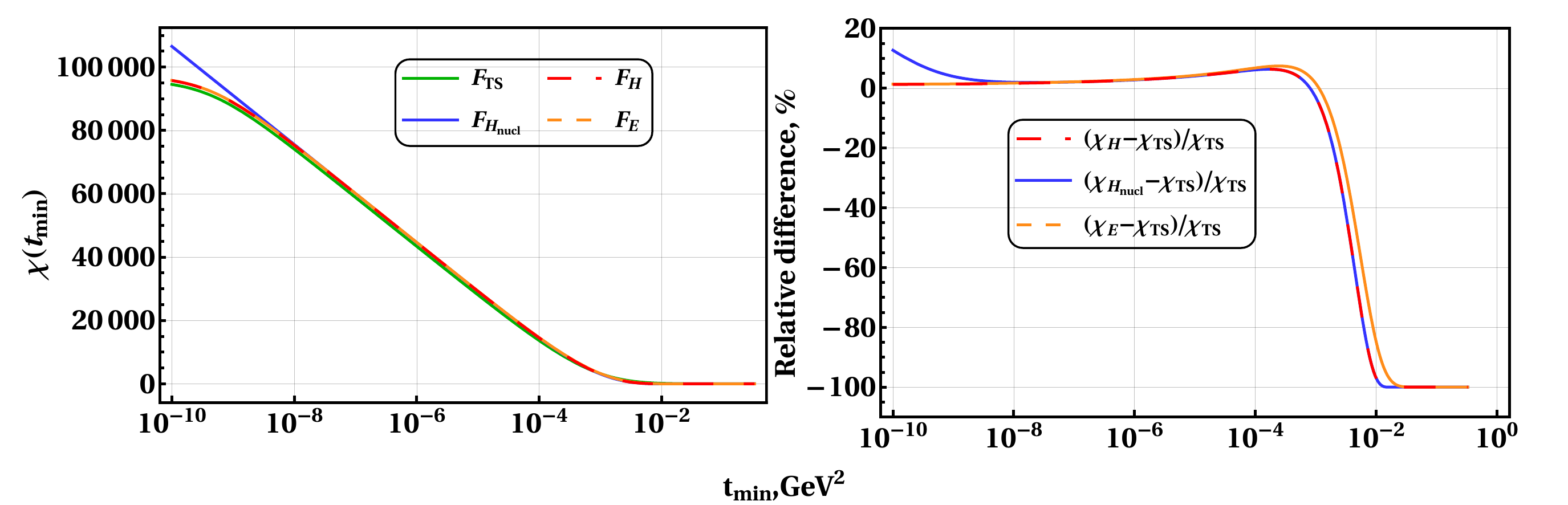}
\caption{Left panel: virtual photon flux for Tsai-Schiff's $F_{TS}$ (the green solid line),  nuclear Helm's $F_{H_{nucl}}$ (the blue solid line), atomic Helm's 
$F_{H}$ (the red dashed line) and atomic exponential $F_{E}$ (the orange dashed line) 
form-factors as function of the lower limits $t_{min}$ with fixed upper 
limit $t_{max} = 10 \; \text{GeV}^2$. Right panel:
the relative difference bewteen benchmark Tsai-Schiff's flux and  other fluxes.
\label{X(tmin,tmax=1)}}
\end{figure*}

Let us consider the kinematic variables of bremsstrahlung-like process $ 2 \to 3$ in the laboratory frame:
\begin{equation}
    l^{-}(p) + N(P_i) \rightarrow l^{-}(p') + N(P_f) + G(k),
\end{equation}
where $p = (E_l, \vect{p}) $ is the momentum of  incoming charged lepton,
$p' = (E'_l, \vect{p'})$ is the momentum of outgoing lepton, $k = (E_G, \vect{k}) $ is the momentum of spin-2 
mediator,  $P_i = (M, 0)$ and  $P_f = (P^0_f,\vect{P}_f)$ are the momenta of the initial and outgoing nucleus 
respectively. We define four-momentum transfer to nucleus as $ P_i - P_f =  q =(q_0, \vect{q})$, such that the 
photon virtuality takes the form $t = -q^2 > 0$.
In order to calculate the differential cross-section of $G$ boson production in nuclear 
interaction one can exploit the   Weizsäcker-Williams approximation, implying that the energy of incoming lepton is much higher than $m_l$ and $m_G$. In this case, the incoming charged particle is replaced by its
effective  photon flux, such that the phase space of the process 
$l^-(p)N(P_i)\to l^-(p')N(P_f)G(k)$ is reduced to the Compton-like process $l^-(p) \gamma(q) \to l^-(p') G(k)$.  In particular,  one can obtain the following expression for the double differential 
cross-section~\cite{Kim:1973he-IWW,Tsai:1973py-PPB,Bjorken:2009-FTE}:
\begin{align}
&     \left.\frac{d \sigma ( p + P_i \rightarrow  p' + P_f + k )}{d(pk)d(kP_{i})}\right|_{WW} = 
\nn  \\   
&    
=    \frac{\alpha \chi}{\pi (p'P_i)}
    \left. \frac{d \sigma ( p + q \rightarrow  k+ p' )}{d(pk)} \right|_{ t = t_{min} },
    \label{DoubleDiffWW1}  
\end{align}
where the flux of virtual photon $\chi$ from nucleus is expressed through the elastic form-factor $F(t)$ as follows:
\begin{equation}
\chi = 
   Z^2 \int\limits^{t_{max}}_{t_{min}} \frac{t - t_{min}}{t^2} F^2(t) dt,
   \label{ChiDefininition1}
\end{equation}
where 
$Z$ is the atomic number of the nucleus, the explicit expressions for the form-factors $F(t)$ are 
discussed  below in Section~\ref{SectionFluxDetails}. 
The photon flux for the inelastic form-factor is proportional to $ \propto Z$, and thus can be ignored in the calculation for heavy nuclei $Z\propto \mathcal{O}(100)$. The WW approach in Eqs.~(\ref{DoubleDiffWW1}) and 
(\ref{ChiDefininition1})  implies that the  virtuality $t$ has its minimum  $t=t_{min}$ when ${\bf q}$ is
collinear with  ${\bf k}- {\bf p}$. In particular,  the expression for $t_{min}$ is derived  below (see, 
e.~g.~Eq.~(\ref{tminDefinition1})).  For ultra-relativistic incident lepton  in laboratory frame we have: 
\begin{align}
      d(k p)d(k \mathcal{P}_i) 
\simeq &
    |\vect{J}(\cos(\theta_{G}),  E_{G}) | d \cos(\theta_{G}) d E_{G} \simeq
\nn     \\
& \simeq   M |\vect{p}| |\vect{k}| d\cos(\theta_{G})d E_{G}, 
    \label{Jacobian1}  
\end{align}
where $\theta_{G}$ is the angle between the initial lepton direction and the momentum of the produced
$G$-boson and $\vect{J}(\cos(\theta_{G}),  E_{G})$ is the Jacobian of the transformation from $(k p)$ and 
$(k \mathcal{P}_i)$ variables to $\cos \theta_G$ and $E_G$. So that by substituting Eq.~(\ref{Jacobian1}) into Eq.~(\ref{DoubleDiffWW1}), we get the following expression  for the double differential cross-section after some algebraic simplification:

\begin{align} 
&     \left.\frac{d \sigma ( p + P_i \rightarrow  p' + P_f + k )}{dx d\cos(\theta_{G})}\right|_{WW}
=  
\nn \\
&     \frac{\alpha \chi}{ \pi }
    \frac{E_l^2 x \beta_{G}}{1-x}
    \left. \frac{d \sigma ( p + q \rightarrow  k + p' )}{d(pk)} \right|_{ t = t_{min} },
\label{dsWW}
\end{align}
 where $\alpha = e^2 / ( 4 \pi ) \simeq 1/137$ is the fine structure constant,  
 $x = E_{G}/E_{l}$ is the energy fraction of spin-2 mediator that it carries
away and $\beta_{G} = \sqrt{1 - m_{G}^2 / (x E_l)^2}$ is the typical velocity of $G$-boson. By solving the  mass-shell equations for both outgoing electron and nucleus:   
\begin{equation}
    p'^2 = (q + p - k)^2 = m_l^2, \quad      P_f^2 = (P_i - q)^2 = M^2,    
    \label{massShellCond1}
\end{equation}
we get the following auxiliary expressions: $q_0 = - t/(2M)$ and $|\vect{q}|^2 = t^2/(4M^2) + t$. 
Furthermore, by taking into account the small typical energy
transferred to the nucleus $|\vect{q}|/M \ll 1$, we get both  approximate expressions $ t \simeq |\vect{q}|^2$ and 
$q_0 \simeq -|\vect{q}|^2/(2M) $  for the  photon  virtuality and nucleus energy transfer respectively.
The value  of $q_0$ is fairly small and thus can be neglected in the calculation,  
since the typical momentum is $|\vect{q}| \lesssim \mathcal{O}(100)\, \mbox{MeV}$ and the  mass of nucleus is of the 
order of  $M \propto \mathcal{O}(100)\,\mbox{GeV}$. On the other hand, the photon flux $\chi$ is sensitive to  the photon 
virtuality   $t \simeq |\vect{q}|^2$, as long as  the screening effects due to the atomic electrons should be taken into account (see e.~g.~Sec.~\ref{SectionFluxDetails} for detail). 

Next, let us introduce the Mandelstam variables in the following form:
\begin{equation}
    s_2 = ( p + q )^2, 
\quad 
    u_2 = ( p - k )^2, 
\quad
    t_2 = (p - p')^2.
    \label{MandelstamDef1}
\end{equation}
It is worth mentioning that  Eq.~(\ref{massShellCond1}) implies, $q_0^2 - |\vect{q}|^2 + u_2 + 2 q_0 V_0 - 2(\vect{q},\vect{V}) = m_{l}^2,$
where we define  the three-vector as $\vect{V} = \vect{p} - \vect{k}$. 
Then for the  small energy transferred to the nucleus $ |\vect{q}|^2 \ll s_2, t_2, u_2 $, one obtains:
\begin{equation}\label{eq_t_q_in_t2}
    t = \frac{(u_2-m_l^2)^2}{4 |\vect{V}|^2\cos^2(\theta_{\vect{q}\vect{V}})},
\end{equation}
where $ \theta_{\vect{q}\vect{V}}$ is the angle between vector $\vect{q}$ and $\vect{V}$. 
Therefore keeping only leading terms in $m_G^2/E_k^2$, $m_l^2/E_l^2$, $m_{l'}^2/E_{l'}^2$ and 
$\theta_{G}^2$, we obtain the approximate expression for the absolute value of the three vector, 
$|\vect{V}|\simeq E_l(1-x)$.
Let us introduce the following auxiliary function:
\begin{equation} 
\label{eq_vect_U}
    U \equiv m_l^2 -u_2 \simeq 
    E_l^2 \theta_G^2 x + m_G^2 (1 - x)/x + m_l^2 x
> 
    0, 
\end{equation}
thus by exploiting Eqs.~(\ref{eq_t_q_in_t2}) and (\ref{eq_vect_U})
one can obtain the expressions for minimum of the virtuality $t = t_{min}$, implying  that $\theta_{\vect{q}\vect{V}} \simeq  \pi$,
such that the three vector $\vect{q}$ is almost collinear with $\vect{k} - \vect{p}$:
\begin{equation}
    t_{min}\simeq  |\vect{q}|^2 \simeq U^2/(4E_l^2 (1-x)^2).
    \label{tminDefinition1}
\end{equation}
It is worth noting that the WW approximation  for the $2\to 3$ cross--section 
 implies that $t_{min}$ is the function of both $x$ and 
 $\theta_{G}$ variables, so that it is fairly accurate approach for the exact tree level 
 cross-section  (for detail, see, e.~g.~Ref.~\cite{Kirpichnikov:2021jev} 
 and references therein).  
Moreover, we note that  $\theta_{\vect{q}\vect{p}} \simeq \pi $  as long as   
$\theta_G \ll 1$ and ${\bf q}$ is 
collinear with  ${\bf k}- {\bf p}$, therefore this yields, $(q, k) \simeq |\vect{q}| |\vect{k}| \simeq Ux/(2(1-x))$,
and we finally get the following expressions for Mandelstam terms  
expressed through the  both  $x$ and $\theta_G$ variables: 
\begin{equation} \label{eq_t2_U}
    u_2 =  m_{l}^2 - U \lesssim  0,
\end{equation}
\begin{equation} 
    t_2 = - 2 (q, k) + t + m_G^2 
\simeq   
    -Ux/(1-x) + m_G^2 
\lesssim     
    0,
\end{equation}
\begin{equation} \label{eq_s2_U}
    s_2 = 2m_l^2 + m_{G}^2 - t - t_2 - u_2 \simeq U/(1 - x) + m_l^2 
\gtrsim  0.
\end{equation}
Note that both the energy conservation law, $q_0+ E_l = E_l' +E_G$, and the condition, $q_0 / E_l  \ll 1$, 
imply that  $E_l \simeq E_{G} + E'_{l}$, thus for the energy faction $x = E_G/E_l$ we get respectively 
its minimum and  maximum  values in the following form  
$\hat{x}_{min} \simeq m_G / E_l$ and $\hat{x}_{max} \simeq  1 - m_l/E_l$. 

By exploiting  FeynCalc package~\cite{Shtabovenko:2016sxi,Shtabovenko:2020gxv} for the Wolfram Mathematica 
routine~\cite{Mathematica},
we get matrix element for  process $l^-\gamma \to l^- G$ (see e.~g.~Appendix~(\ref{MatElDeriv})): 
\begin{widetext}
\begin{multline}
     \left| \mathcal{M}_{l^{-} \gamma \rightarrow G l^{-}} \right|^2 
= 
    \frac{c_l^2 e^2}{\Lambda^2}
    \frac{u_2 [s_2 - 2 m_l^2] [(t_2 + u_2)^2 + (u_2 - m_G^2)^2] [4 u_2 (2 m_l^2 - s_2) -  m_G^2 t_2] }
         {4 t_2 (u_2 - m_l^2)^2 (s_2 - m_l^2)^2}
-  \\ - 
    \frac{c_l^2 e^2}{\Lambda^2}
    \frac{m_l^2 R(m_l,m_G, t_2,u_2)}
         {12 t_2^2 (m_l^2 - u_2)^2 (s_2 - m_l^2)^2 }, 
         \label{MegToGe22}
\end{multline}
\end{widetext}
where  $R(m_l,m_G, t_2,u_2)$ is regular expression for $m_l$ and $m_G$ 
that is given in the Appendix~\ref{MatElDeriv}. Note that in 
Eq.~(\ref{MegToGe22}) we set universal coupling of $G$ boson to both charged lepton and 
photon, such  that $c_l \equiv c_e=c_\mu = c_{\gamma}$. It implies that the unitarity  of the scenario is not violated at
low  energies as soon as $m_{G} \to 0$ (see, e.~g.~Ref.~\cite{Kang:2020-LGMDM} and references therein for detail). 
The differential cross section for process $2 \to 2$ is 
\begin{equation}\label{8pissMM}
    \frac{d \sigma_{2\to 2}}{d (p k)}
= 
    2 \frac{ d\sigma_{2 \to 2}}{d u_2} 
=
    \frac{1}{8 \pi (s_2 - m_l^2)^2} \left| \mathcal{M}_{l^{-} \gamma \rightarrow  G l^{-}} \right|^2,
\end{equation}
where $\left| \mathcal{M}_{l^{-} \gamma \rightarrow  G l^{-}} \right|^2$ 
is defined by (\ref{MegToGe22}). As a result, the double-differential cross section of the  bremsstrahlung-like process 
$l N \to l N G (\to \chi \overline{\chi})$  takes the following form:

\begin{align}
&
    \left.\frac{d \sigma ( p + P_i \rightarrow  p' + P_f + k )}{dx d\cos(\theta_{G})}\right|_{WW}
=
\nn \\
&    =
    \frac{\alpha \chi}{ \pi }
    \frac{E_l^2 x \beta_{G}}{1-x}
    \frac{1}{8 \pi (s_2 - m_l^2)^2}
    \left| \mathcal{M}_{l^{-} \gamma \rightarrow  G l^{-}} \right|^2.
    \label{eq:dsWWRes}
\end{align}

In order to verify our calculation for the matrix element 
squared of spin-2 particle production  $e^-(p) \gamma(q) \to G(k) e^-(p') $ one can exploit the crossing 
symmetry for the well known process of electron positron pair 
annihilation $ e^+(p_{e^+}) e^-(p_{e^-}) \to G(p_G) \gamma (p_{\gamma}') $, for which the amplitude squared  can be found in Ref.~\cite{Kang:2020-LGMDM}.  In 
particular,  for the massless fermion the regarding transition element  
reads as follows~\cite{Kang:2020-LGMDM}:
\begin{align}
&  \left| \mathcal{M}_{e^{+}e^{-} \rightarrow \gamma G} \right|^2 =
\nn \\
& =  \frac{c_e^2 e^2  }{\Lambda^2} 
    \frac{  \left( \hat{s}_{2}^2 + 2 \hat{t}_{2} (\hat{s}_{2} + \hat{t}_{2}) - 2m_G^2 \hat{t}_{2} + m_G^4 \right) }{4 \hat{t}_{2} \hat{s}_{2}(\hat{s}_{2} + \hat{t}_{2} - m_G^2)}  \times
\nn  \\
& \times \left( 4 \hat{t}_{2} (\hat{s}_{2} + \hat{t}_{2}) - m_G^2 (\hat{s}_{2} + 4 \hat{t}_{2}) \right), 
 \label{MeeToGammaG1}
\end{align}
where the Mandelstam variables are defined by: 
\begin{equation}
    \hat{s}_{2} 
\equiv 
    ( p_{ e^{-} } + p_{e^{+}} )^2, \quad 
        \hat{t}_{2} 
\equiv 
    (p_{ e^{-} } - p'_{G} )^2, \quad 
    \hat{u}_{2} 
\equiv 
    (p_{ e^{-} } - p'_{\gamma}  )^2.     
\end{equation}
    The crossing-symmetry implies the  momentum 
replacement for the initial  process $e^{+}e^{-} \rightarrow \gamma G$ in the following form: 
$    p_{e^{+}} \to - p'_{e^{-}}$ and  $p'_{\gamma} \to - p_{\gamma}$. The regarding  
 Mandelstam variables transform as follows:
$\hat{s}_{2}  \rightarrow t_2$, 
$ \hat{t}_{2}  \rightarrow u_2$ and 
$ \hat{u}_{2}  \rightarrow s_2$,
where the notations for $s_2, u_2$ and $t_2$ are introduced (see e.~g.~Eq.~(\ref{MandelstamDef1}))
in order to match with conventional  labels of the  authors of Ref.~\cite{Bjorken:2009-FTE}. Finally this yields 
the following expression for  the matrix  element squared: 
\begin{align}
&   \left| \mathcal{M}_{e^{-} \gamma \rightarrow  G e^{-}} \right|^2 
=
\nn \\ 
& =   \frac{c_e^2 e^2  }{\Lambda^2}  
    \frac{  \left( t_2^2 + 2 u_2 (u_2 + t_2) - 2m_G^2 u_2 + m_G^4 \right) 
         }{
            4 u_2 t_2 ( u_2+t_2-m_G^2 )} \times 
\nn \\
& \times \left( 4 u_2 (u_2 + t_2) - m_G^2 (t_2 + 4 u_2) \right)
  \label{MeGammaToeG1}
\end{align}
%
It is worth noting that Eq.~(\ref{MegToGe22})  tends to  Eq.~(\ref{MeGammaToeG1}) in the 
massless lepton limit as soon as  $m_l \to 0$.

\section{The virtual photon flux function
\label{SectionFluxDetails}}

  \begin{figure*}[!tbh]
    \centering
    \includegraphics[width=0.65\textwidth]{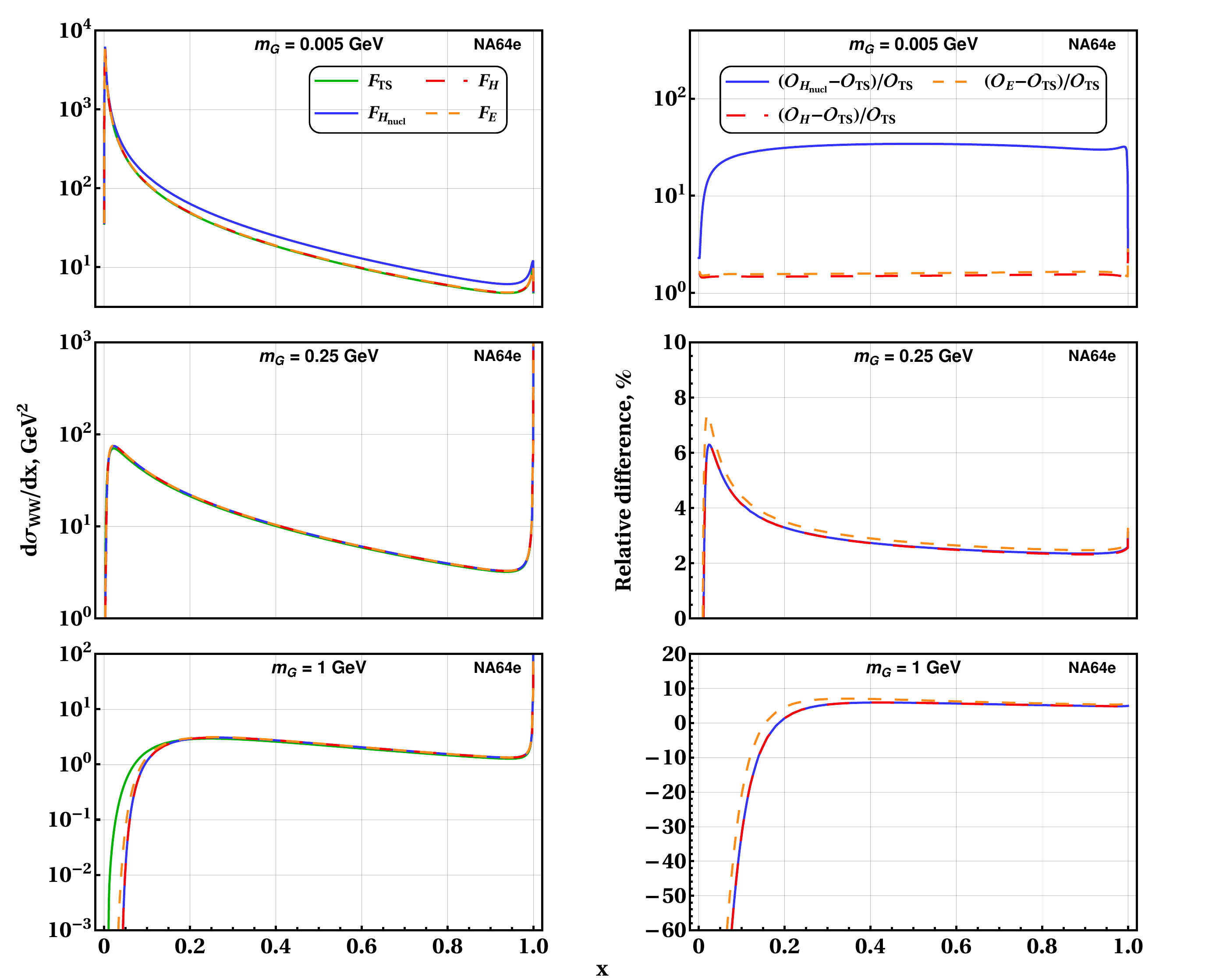}
    \caption{Left panel: the differential cross-section of a $G$-boson production at NA64$e$ experiment as function
    of the energy fraction for the set of its mass and the set of form-factors. 
    The set of masses is chosen to be: $m_{G}=5\, \mbox{MeV}$, $m_{G}=  250\, \mbox{MeV}$ and $m_{G}= 1\, \mbox{GeV}$. 
    The cross-section for $F_{TS}$  is 
    shown by the green solid line, the cross-section for $F_{H_{nucl}}$  is shown by blue solid line,
    the cross-section for $F_{H_{nucl}}$  is shown by blue solid line,
    the cross-section for $F_{H}$  is shown by the red dashed line, the cross-section for $F_{E}$  is shown by
    the orange dashed line. Right panel: the  regarding relative 
    differences of the the cross-sections for various form-factor parametrizations are expressed as 
    $(\mathcal{O}_{H_{nucl}} - \mathcal{O}_{TS})/\mathcal{O}_{TS}$, 
    $(\mathcal{O}_{H} - \mathcal{O}_{TS})/\mathcal{O}_{TS}$ and $(\mathcal{O}_{E} - \mathcal{O}_{TS})/\mathcal{O}_{TS}$ 
    respectively, where $\mathcal{O}\equiv d\sigma/dx$. 
    \label{fig:dsdx_NA64e} }
\end{figure*}

In this section we discuss the impact of  different atomic and nuclear form-factor parametrization
$F(t)$ for  the photon flux $\chi$ which is given by Eq.~(\ref{ChiDefininition1}). The latter 
affects  the  differential and total cross sections of hidden spin-2 boson production, that will be  discussed in Section~\ref{CS_WW}. 

The nuclear form-factor in the laboratory frame is associated with charge density of nucleus through the Fourier transformation
for both spin-$0$ and  spin-$1/2$ (see e.~g.~Refs.~\cite{PERDRISAT2007694,RevModPhys.35.335,Bjorken:2009-FTE,PhysRevD.95.036010,PhysRevD.96.016004,Kirpichnikov:2021jev,Tsai:1973py-PPB,DRELL196418,RevModPhys.35.335} and references therein for detail).  
The atomic form-factor can be represented as the nuclear form-factor that takes into account 
the screening of the nucleus by  Coulomb field due to the atomic electrons. Indeed, in the limit 
$t \to 0$  nuclear form-factor tends to $F_{nucl}(t)\to 1$ in opposite to the atomic form-factor, 
which tends to $F_{atom}(t)\to 0$ as $t\to 0$.
In addition, screening charge density of the atomic form-factor can be represented as a convolution of the 
nuclear charge density with the specific screening density. 
In particular, following L.~Schiff~\cite{Schiff:1951elFF} to obtain the atomic form-factor one should 
multiply the nuclear form-factor by the screening  term $t/(t_a + t)$.

First, let us consider the elastic atomic form-factor that was studied by Y.~Tsai~\cite{Tsai:1973py-PPB} and L.~Schiff~\cite{PhysRev.92.988} in the following form~\cite{Bjorken:2009-FTE}:
\begin{equation}
    F_{TS}(t)  = 
    \frac{t}{(t_a+t)}  \frac{1}{(1 + t/t_d)},
    \label{TsaiFFdefinition11}
\end{equation}
where $\sqrt{t_{a}} = 1/R_a$ is a momentum transfer associated with nucleus Coulomb field screening due to the atomic
electrons, with $R_a$ being a  typical  magnitude of the atomic radius $R_a = 111 Z^{-1/3}/m_e$, 
$\sqrt{t_{d}} = 1/R_n$ is the typical momentum associated 
with nuclear radius $R_n$, such that $R_n\simeq 1/\sqrt{d}$ and $d = 0.164 A^{-2/3}  \text{GeV}^2$.
Since the integration with respect to $t$ in Eq.~(\ref{ChiDefininition1})  is dominated by $t\gtrsim t_{min}$, 
therefore the magnitude of $t_{min}$ (see e.~g.~Eq.~(\ref{tminDefinition1})) defines the typical form-factor 
approach to be considered.   In particular, if  $t_{min}/t_a \ll 1$, then the complete screening regime takes 
place,  which implies that nucleus  transfer momentum is small and the typical atomic elastic form-factor is much 
less then unity, $F_{TS}(t) \simeq t/t_a \ll 1$.     
On the other hand, as soon as $t_{min}/t_a \gg 1$, then no screening regime occurs. In this case the atomic elastic 
form-factor is scaled as $F_{TS} (t) \simeq 1/(1+t/t_d)$ and the nucleus size effects dominate, it implies also that 
the typical   nucleus transfer momentum squared is relatively large, $t_d \gg t_a$. 
It is worth noting,  given the  parameter space of interest 
$1\, \mbox{MeV} \lesssim m_G\lesssim 1\, \mbox{GeV}$ 
and  $10 \, \mbox{GeV} \lesssim  E_l \lesssim  100\, \mbox{GeV}$  both  
screening and nucleus size parameters can contribute to the virtual photon flux and thus to the total 
yield  of the $G$-boson production.  

Next, let us consider the nuclear Helm's form-factor $F_{H_{nucl}}(t)$ 
that corresponds to the inverse Fourier transformation  of the nucleus charge density $\rho(\vec{r})$. The latter can be 
represented  as the  convolution of the spherically uniform charge inside the nucleus and the Gaussian profile implying better 
accounting  of the the edge of nucleus~\cite{Chen:2011xp}. Both the nuclear Helm's form-factor $F_{H_{nucl}}(t)$ and atomic 
Helm's form-factor $F_{H}(t)$  read as follows respectively ~\cite{Lewin:1995rx-RDM,Dobrich:2015jyk-ALPt}:
\begin{align}
&    F_{H_{nucl}}(t) 
= 
   \frac{3 j_1( \sqrt{t} R_{H} )}{\sqrt{t} R_{H}} 
   \exp{- s_{H}^2 t / 2},
\\
& F_{H}(t) = t/(t_a+t) F_{H_{nucl}}(t),
\end{align}
where $j_1(x) $ - first spherical Bessel function of the first kind, the  
effective nuclear radius $R_H$ can be parameterized as
\[
    R_{H} = \sqrt{c_{H}^2 + 7/3 \pi^2 a_{H}^2 - 5 s_{H}^2},
\]
where $s_{H} = 0.9 \; \text{fm}$ is the nuclear shell thickness,  $ a_{H} = 0.52 \; \text{fm}$  and $c_{H} = (1.23 A^{1/3} - 0.6) \; \text{fm}$.  
It is worth noting that we also set $F_{H_{nucl}}$ to zero for $t \gtrsim (4.49/R_{H})^2$, it implies that we 
consider only non-negative values of the Helm's form-factor, since it vanishes at $t \simeq
(4.49/R_{H})^2$ and the dominant contribution for the photon flux is due to the typical 
range $t \lesssim (4.49/R_{H})^2$  (for details see  e.~g.~Ref.~\cite{Dobrich:2015jyk-ALPt}).

Finlay, let us consider now the exponential atomic form-factor corresponding Gaussian 
charge distribution that reads as follows \cite{PhysRevD.37.3388}:
\begin{equation}
    F_{E}(t) =  t/(t_a+t) \exp\left( - t R^2_{exp} /6 \right),
    \label{ScreeninAtomExpFF}
\end{equation}
where the mean radius of nucleus  is defined as
$R_{exp} = (0.91 A^{1/3} + 0.3) \text{fm}$. As was discussed above, the screening term $t/(t_a+t)$ in 
Eq.~(\ref{ScreeninAtomExpFF}) is introduced to take into account  the shielding of the 
nucleus Coulumb field due to the atomic electrons.

Note that the general virtual photon flux (\ref{ChiDefininition1}) depends on both fraction of energy $x$ and  emission 
angle  of spin-2  boson  $\theta_G$ through the  function of the lower limit $t_{min}(x,\theta_G)$
(see e.~g.~Eq.~(\ref{tminDefinition1}) for detail). So that it is instructive to study the the  dependence of $\chi$ 
upon $t_{min}$ for various form-factor  parametrizations. 
Intriguingly, that the integration over $dt$ in the virtual photon flux (\ref{ChiDefininition1})  
of  the Tsai-Schiff's elastic form-factor  (\ref{TsaiFFdefinition11}) can be performed in the 
analytical way through the  elementary  functions. The latter simplification can be exploited
for the  reducing of the computational time of $\chi_{_{TS}}$ integration.  
As the result, the  virtual photon flux in analytic form for Tsai-Schiff's form-factor  reads  as follows:
\begin{align}
&  \chi_{_{TS}} 
=    \frac{Z^2 t_d^2}{(t_a - t_d)^3}
 \Bigl(   	    \left[
    	        C^{\chi}_{1}
    	    + 
    	        C^{\chi}_{2} t_{min}
    	    \right]
    	+
\nn \\
&
	+
    	    \left[
    	            C^{\chi}_{3}
    	        +   C^{\chi}_{4} t_{min} 
    	    \right]
    	    \ln{\left[ \frac{t_{min} + t_d}{t_{min} + t_a}\right] }
    	    \Bigr) 
    	\label{TsaiFF}
\end{align}
where the functions $C_1^\chi$, $ C_2^\chi$,  $C_3^\chi$ and  $C_4^\chi$ are defined by the following  expressions respectively:
\begin{align}
&    		C^{\chi}_{1} 
	=
        \Bigl(
		 	\frac{t_d(t_a - t_d)}{t_d + t_{max}}
		+	\frac{t_a(t_a - t_d)}{t_a + t_{max}}
		- 	
 \\   
 &   
 	- 	
		    2 (t_a - t_d)
		+	(t_a + t_d) 
		\ln{\left[ \frac{(t_d + t_{max})}{(t_a + t_{max})}\right] }
		\Bigr),
\end{align}
\begin{equation}
		C^{\chi}_{2}
	=
		\left( 
		\frac{t_a - t_d}{t_d + t_{max}}
		+	\frac{t_a - t_d}{t_a + t_{max}}
		+	2 \ln{\left[ \frac{t_d + t_{max}}{t_a + t_{max}}\right] }
		\right),     
\end{equation}
\begin{equation}
    				C^{\chi}_{3}
	=
	-	(t_a + t_d),
	\quad
		C^{\chi}_{4}
	=
	-	2.
\end{equation}

The regarding flux for 	and  exponential form-factor can be expressed through the special function as follows
\begin{align}
    &     \chi_{_{E }}
=  Z^2
\Bigl(
    \frac{t_a + t_{min}}{t + t_a}
    e^{- R_{exp}^2 t / 3 } 
 +   
  \label{ScrExpFF}   \\
    & +   \frac{3 + R_{exp}^2 (t_a + t_{min})}{3}
    e^{R_{exp}^2 t_a/3}
    \text{Ei}\left[ - \frac{R_{exp}^2 (t + t_a)}{3} \right]
\Bigr) \Bigr|_{t_{min}}^{t_{max}},  
   \nn 
\end{align}
where  $\text{Ei}(x) = \int_{- \infty}^{x} e^t / t \; dt$ is the exponential integral
function.  Contrary, photon flux with Helm's form-factor cannot be integrated over $t$ in analytical way, in what 
follows  we  perform the numerical integration for both $\chi_{_{H}}$ and $\chi_{_{H_{nucl}}}$. 
In addition, the numerical calculations reveal that general photon flux (\ref{ChiDefininition1}) 
depends weakly on $t_{max}$ as long as $t_{max} \gg t_{min}, t_a, t_d$.  
On the other hand, one can show that the typical values of the $t_{min}, t_a$  and $t_d $ don't 
exceed the magnitude of the order of  $ \lesssim \mathcal{O}(1)\, \mbox{GeV}^2$.  In what follows  
we set  $t_{max} \simeq 10\, \mbox{GeV}^2$ for the numerical estimations.

In the left panel of Fig.~\ref{X(tmin,tmax=1)} we show the  virtual photon fluxes $\chi$ as a functions of 
$t_{min}$ for the  lead target material of the NA64$e$ experiment with atomic number of $Z=82$ and particle 
number of $A=207$ that parameterize both typical screening and  nuclear radius. 
In the right panel of Fig.~\ref{X(tmin,tmax=1)} we choose $\chi_{_{TS}}$ as a benchmark photon 
flux and show the the relative differences between it and the set of $\chi_{_{E}}$, 
$\chi_{_{H}}$ and $\chi_{_{H_{nucl}}}$.  For the   wide range of the typical momentum squared
$10^{-10}\, \mbox{GeV}^2 \lesssim  t_{min} \lesssim 10^{-3}\, \mbox{GeV}^2$  all the atomic 
photon fluxes match with a reasonable accuracy at the level of $ \lesssim 5 \%$. 
In order to illustrate the impact of the screening effect 
we show both the nuclear and atomic Helm's photon fluxes in Fig.~\ref{X(tmin,tmax=1)}. 
In particular, for the screening region $t_{min} \lesssim t_a \simeq 10^{-9}\, \mbox{GeV}^2$ the nucleus photon flux 
$\chi_{_{H_{nucl}}}$  exceeds the atomic one $\chi_{_{H}}$  by  $10 \%$ approximately.  The latter one 
is associated also with the smaller slope of $\chi_{_{H}}$ in the screening range~$t_{min} \lesssim t_a$.

It is worth noting,
that there are also form-factor models that consider the charge density 
as the sum of Gaussian,  Fourier-Bessel 
functions and Klein-Nystrand's charge density~\cite{Hofstadter:1956qs,SICK1974509, Dreher:1974pqw,Duda:2006uk,PhysRevC.60.014903}. In addition,  the 
Fermi  distribution  for the nucleus charge is also discussed in the literature~\cite{Lewin:1995rx-RDM,Duda:2006uk}.
We note that these nucleus form-factor parametrizations  are  beyond the scope of the  present 
paper, even though we would expect that they provide a similar results for the flux shown in 
Fig.~\ref{X(tmin,tmax=1)}.

\section{Missing energy signal \label{ExperimentalBenchmark}} 

In this section we discuss the  setups for the 
fixed target experiments such as NA64e (SPS, CERN) 
\cite{Gninenko:2016kpg,NA64:2016oww,NA64:2017vtt,Gninenko:2018ter,Banerjee:2019pds,Dusaev:2020gxi,Andreev:2021fzd,PhysRevLett.126.211802,Andreev:2022hxz,NA64:2022rme,Arefyeva:2022eba,Zhevlakov:2022vio,NA64:2021acr,NA64:2021xzo}, LDMX (Fermilab)~\cite{PhysRevD.99.075001,Ankowski:2019mfd,Schuster:2021mlr,Akesson:2022vza}, NA64$\mu$  (SPS, CERN) 
\cite{PhysRevD.105.052006,Kirpichnikov:2021jev} and M$^3$ (Fermilab) \cite{Capdevilla:2021kcf,Kahn:2018cqs}, that can potentially probe invisible 
decay of $G\to \chi \overline{\chi}$ in the associated charged lepton missing energy process 
$l N \to l N G (\to \chi \overline{\chi})$, where the label $l=(e,\mu)$ denotes either electron or muon beam and $N$
denotes the nucleus of the target. Given the benchmark  coupling
Eq.~(\ref{BechmarkLagrOftheModel1}), the spin-2 Dark Matter mediator $G$ can decay through 
the different  channels. In particular, as soon as  $m_G \gtrsim 2 m_l$ 
the visible decay $G\to l^+ l^-$ is allowed with a specific decay width~\cite{lee:2014GMDM} 
\begin{align}
    &      \Gamma(G \to l^+ l^-)
=
    \frac{c_l^2 m_G}{160 \pi}
    \left( \frac{m_G}{\Lambda} \right)^2 \times 
\nn     \\
    & \times 
     \left( 1 - 4 m_l^2/m_G^2 \right)^{3/2}
    \left( 1 + 8 m_l^2/(3m_G^2) \right),
    \label{GtollDecayWidth}
\end{align}
where $m_l$ is the mass of the charged lepton. The Lagrangian (\ref{BechmarkLagrOftheModel1})
also implies that for $m_{G} \gtrsim 2 m_\chi$ the invisible decay into fermion DM pair 
$G\to \chi \bar{\chi}$ is kinetically allowed with a decay width 
\begin{align}
    &     \Gamma(G \to \chi \bar{\chi})
=
    \frac{c_\chi^2 m_G}{160 \pi}
    \left( \frac{m_G}{\Lambda} \right)^2 \times 
\nn     \\ 
    & \times    \left( 1 - 4 m_\chi^2/m_G^2 \right)^{3/2}
    \left( 1 +   8 m_\chi^2/(3 m_G^2) \right),
\label{Gto2chiDecayWidth}    
\end{align}
where $m_\chi$ is the mass of the hidden Dirac DM fermion.
In this paper we focus on the processes of the invisible channel of $G$-boson decay into pair of hidden
dark fermions, $G\to \chi \bar{\chi}$ with $\mbox{Br}(G\to \chi \bar{\chi})\simeq 1$ for the 
sufficiently light DM particles, $m_G \gtrsim 2m_\chi$.  It implies that decay widths obey the condition 
$\Gamma_{G\to \chi \bar{\chi}} \gg \Gamma_{G \to l^+ l^-}$ and therefore the coupling constants are chosen to be  
$c_{\chi} \gg c_l $ for the parameter space of interest $1\, \mbox{MeV} \lesssim  m_{G} \lesssim 1\, \mbox{GeV}$. 
As the result, this benchmark conditions imply the rapid decay of spin-2 DM 
mediator to $\chi \bar{\chi}$ pair after its production in the process $l N \to l N G$. 

\begin{figure*}[!tbh]
    \centering
    \includegraphics[width=0.65\textwidth]{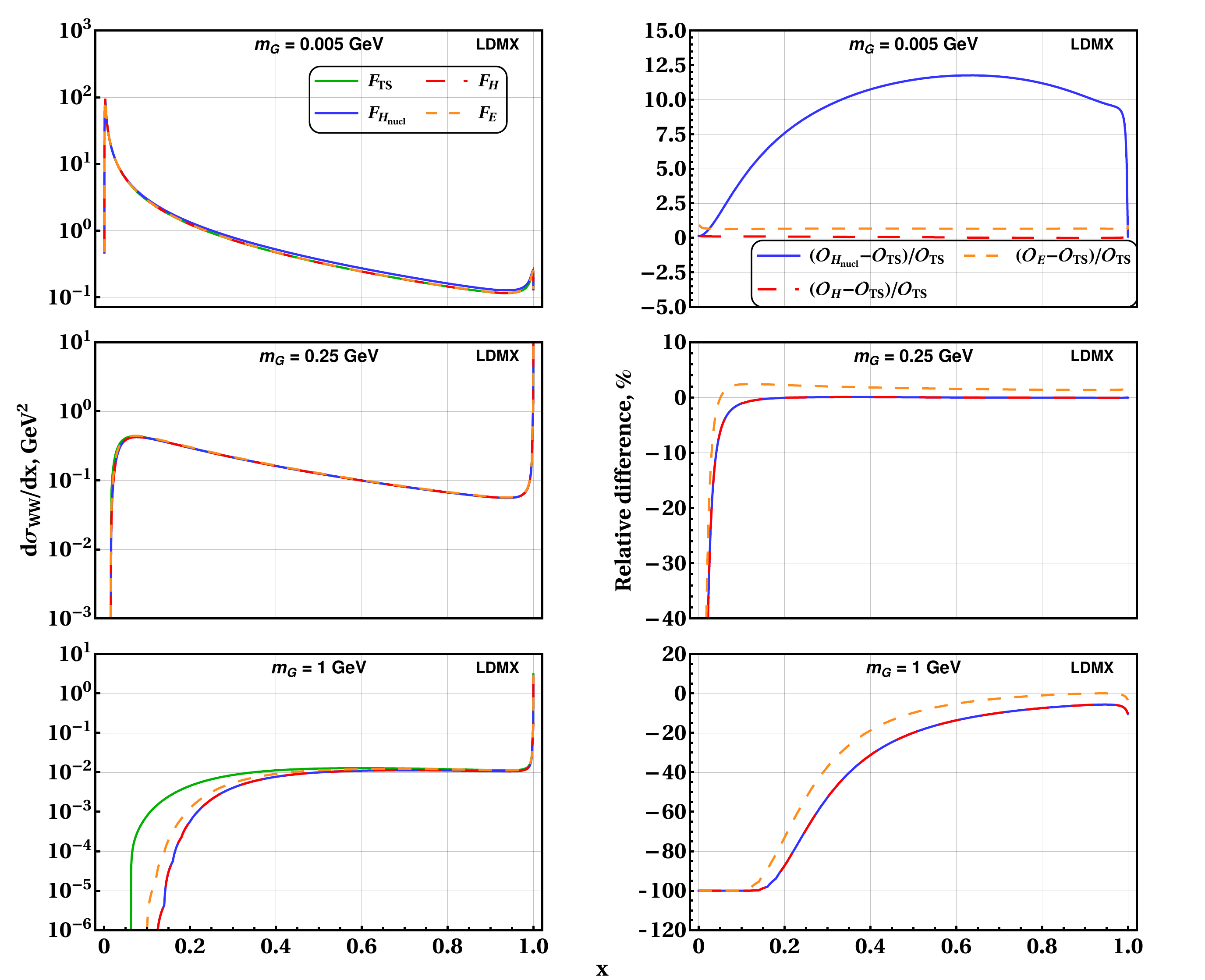}
    \caption{Same as in Fig.~\ref{fig:dsdx_NA64e} but for the LDMX experiment.
    \label{fig:dsdx_LDMX} }
\end{figure*}

Let us estimate $N_{G}$ the number of  $G$ produced by the lepton beam  at fixed target as follows
 \begin{equation}\label{eq:NradGrav}
N_{G} \simeq \mbox{LOT}\cdot \frac{\rho N_A}{A} L_T \int\limits^{x_{max}}_{x_{min}}
dx \frac{d \sigma_{2\to3}(E_l)}{dx} \mbox{Br}(G \to \chi \bar{\chi} )\,,
 \end{equation}
where $\mbox{LOT}$ is number of  leptons accumulated on target, $\rho$ is target density,  
$N_A$ is Avogadro's number,  $A$ is atomic weight number,  
$L_T$ is the effective interaction length of the lepton in the target, $d\sigma_{2\to3}/dx$ is the differential cross-section of the lepton missing energy process $l N \to l N G(\to \chi \bar{\chi})$, 
$E_l$ is the initial energy of lepton beam,   $x_{min}$ and $x_{max}$
are the minimal and maximal fraction of missing energy respectively
for the  regarding  experimental setup, $x\equiv E_{miss}/E_l$, where $E_{miss} \equiv  E_G$. 
The $x_{min} \lesssim x \lesssim x_{max}$ cuts are determined by specific fixed–target facility.  
In what follows, we describe below the input benchmark parameters of the lepton
fixed-target experiments.

\begin{figure*}[!tbh]
    \centering
    \includegraphics[width=0.65\textwidth]{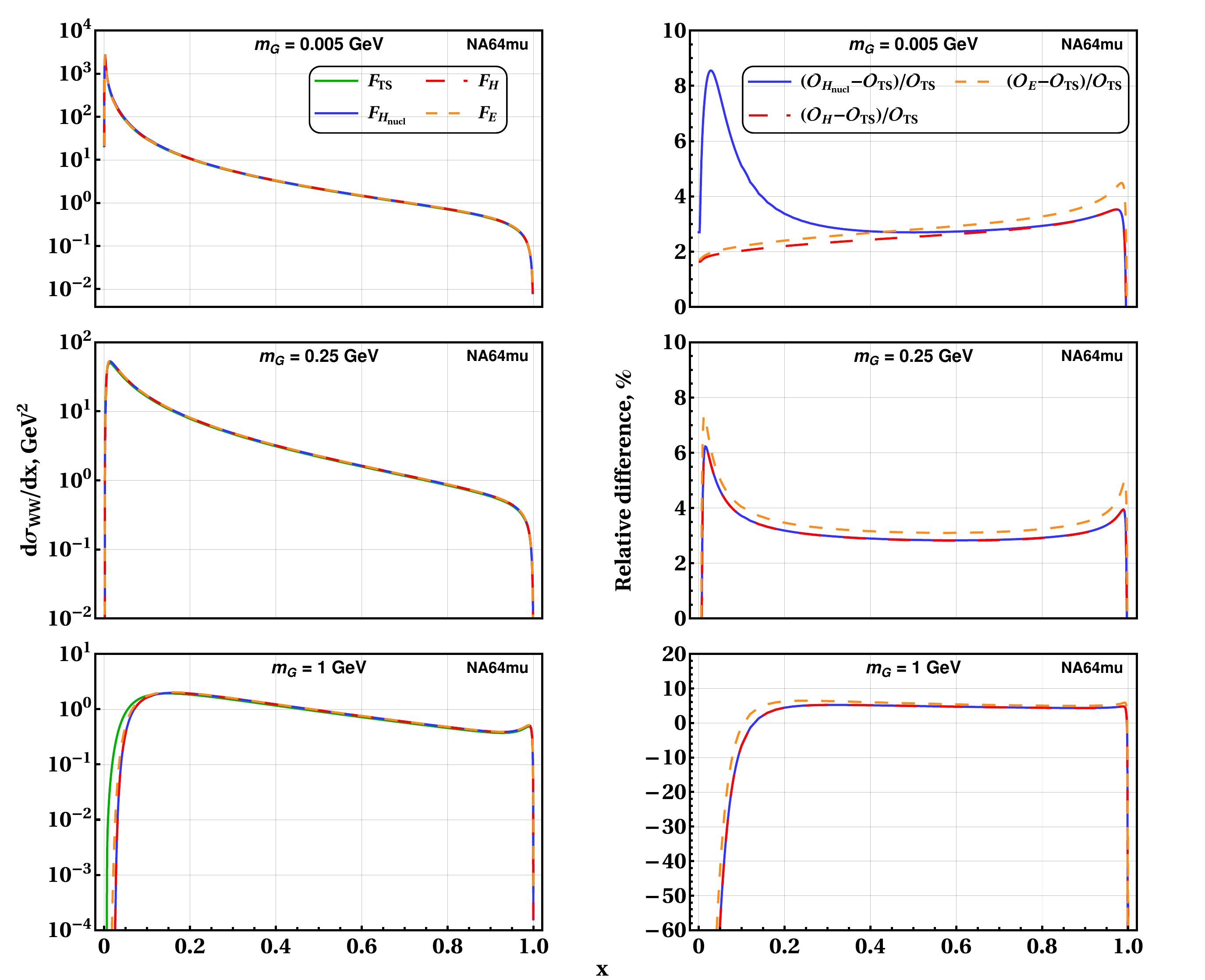}
    \caption{Same as in Fig.~\ref{fig:dsdx_NA64e} but for the NA64$\mu$ experiment.
    \label{fig:dsdx_NA64mu} }
\end{figure*}

\subsection{NA64$e$
\label{NA64eSetupSect}}

The spin-2 mediator of DM can be produced in the reaction of ultra-relativistiс electrons of 
$E_e\simeq 100\, \mbox{GeV}$ scattering off the nuclei of an active target $e N \to e N G$ followed 
by rapid $G\to \chi \bar{\chi}$ decay into DM particles. The fraction of the primary electron energy $E_{miss}=x E_e$
can be  carried away by  $\chi \bar{\chi}$ pair, that passes the NA64$e$ detector without energy deposition.
The remaining part of the beam energy fraction, $E_e^{rec} \simeq (1-x) E_e$, can be deposited in the electromagnetic 
calorimeter (ECAL) of NA64$e$ by the recoil electrons.  So that, the production of the hidden spin-2 boson can be 
observed as an excess of events with a single electromagnetic shower of energy $E^{rec}_e$ above the predicted  
background~\cite{NA64:2016oww}. In this paper we carry out an estimate that implies the localization of
that electromagnetic shower  in the first radiation length of the lead target. 
As a result one can set  $L_T \simeq X_0$ in Eq.~(\ref{eq:NradGrav}), where $X_0\simeq 0.56\, \mbox{cm}$ is the 
typical radiation length of the electron in the lead target. In addition, the candidate event is required to have 
a missing energy in the range $E^{rec}_e \lesssim 0.5 E_e \simeq 50\, \mbox{GeV}$, that leads to the specific  
 energy fraction cut $x_{min}\simeq 0.5$ for the NA64$e$ facility in Eq.~(\ref{eq:NradGrav}). 

 Moreover we note that the 
 electromagnetic calorimeter of the NA64$e$ serves as the active target for the incident electron beam and contains 
 $6\times 6$ Shahalyk-type modules which are made of the both plastic scintillator (Sc) and  lead  (Pb,  
 $\rho \simeq 11.34\, \mbox{g cm}^{-3}$, $A=207\, \mbox{g mole}^{-1}$, $Z=82$) plates.  It is worth mentioning 
 that the production of $G$-boson inside the scintillator plate is subdominant due to its smaller density, 
 $\rho(\mbox{Sc}) \ll  \rho(\mbox{Pb})$, and larger radiation length, $X_0(\mbox{Sc}) \gg X_0(\mbox{Pb})$. In what follows  in
 the numerical  calculations we neglect the contribution of Sc to the  production rate of the spin-2  mediator.  
 
To conclude this subsection we note that NA64$e$ uses the electron beam from  
 the $H4$ beam line at the CERN SPS. The regarding beam intensity is estimated to be of the order of $\simeq 10^7$ 
 electrons per spill of $\simeq 4.6\, \mbox{seconds}$,
 however the number of good spills per day is expected to be $\simeq 4000$. As 
 a result, about $\simeq 120$ days are needed to collect $\mbox{EOT} \simeq 5\times 10^{12}$ at the $H4$ beam line for the projected
 statistics of the NA64$e$. In this work we also perform the analysis of the sensitivity of NA64$e$ to probe DM for the 
 $\mbox{EOT} \simeq 3.22 \times 10^{11}$ (see e.~g.~Ref.~\cite{Andreev:2022hxz} for detail).
 The regarding statistics has been already collected by NA64$e$ during previous experimental runs in 2016-2021.

\subsection{LDMX
\label{LDMXsubSect}}

The Light Dark Matter Experiment (LDMX) is the projected electron fixed-target facility at Fermilab, that aims 
probing  the relic DM particles in the mass range between $1\, \mbox{MeV}$ and $1\, \mbox{GeV}$. It can be 
considered as a facility that  is complimentary to NA64$e$ experiment due to its unique electron missing momentum
technique~\cite{Mans:2017vej}.  It is remarkable that the missing-energy cuts and active 
veto system of both NA64$e$ and LDMX experiments provide a significant background 
suppression at the level of $\lesssim 10^{-12}$.

The projected LDMX experiment would employ the target, the silicon tracker and 
both the hadron and electromagnetic calorimeter. The lost of the primary electron beam 
energy can be associated with $G$-boson emission in the thin upstream target of the LDMX. The 
regarding missing momentum of the electron can be measured by the silicon tracker system, 
electromagnetic and hadron calorimeter which are located downstream. The missing energy cut of the recoil electron 
is set to be $E^{rec}_e \lesssim 0.3 E_e$, that implies $x_{min} = 0.7$ in Eq.~(\ref{eq:NradGrav}). We perform the 
analysis of the LDMX sensitivity for aluminium target (Al) $(\rho=2.7\,\mbox{g cm}^{-3}, A=27\,\mbox{g mole}^{-1},
Z=13)$ with a thickness of  $L_T \simeq 0.4 X_0 \simeq 3.56 \,\mbox{cm}$, where $X_0=8.9~\mbox{cm}$
is a radiation length of the electron in the aluminium. The energy of the beam is chosen to be 
$E_e\simeq 16\,\mbox{GeV}$ and the projected statistics corresponds to $\mbox{EOT}\simeq 10^{16}$ for the final phase of
experimental running after 2027 (see e.~g.~Ref.~\cite{Akesson:2022vza} and references therein for detail).

\subsection{NA64$\mu$}

\begin{figure*}[!tbh]
    \centering
    \includegraphics[width=0.65\textwidth]{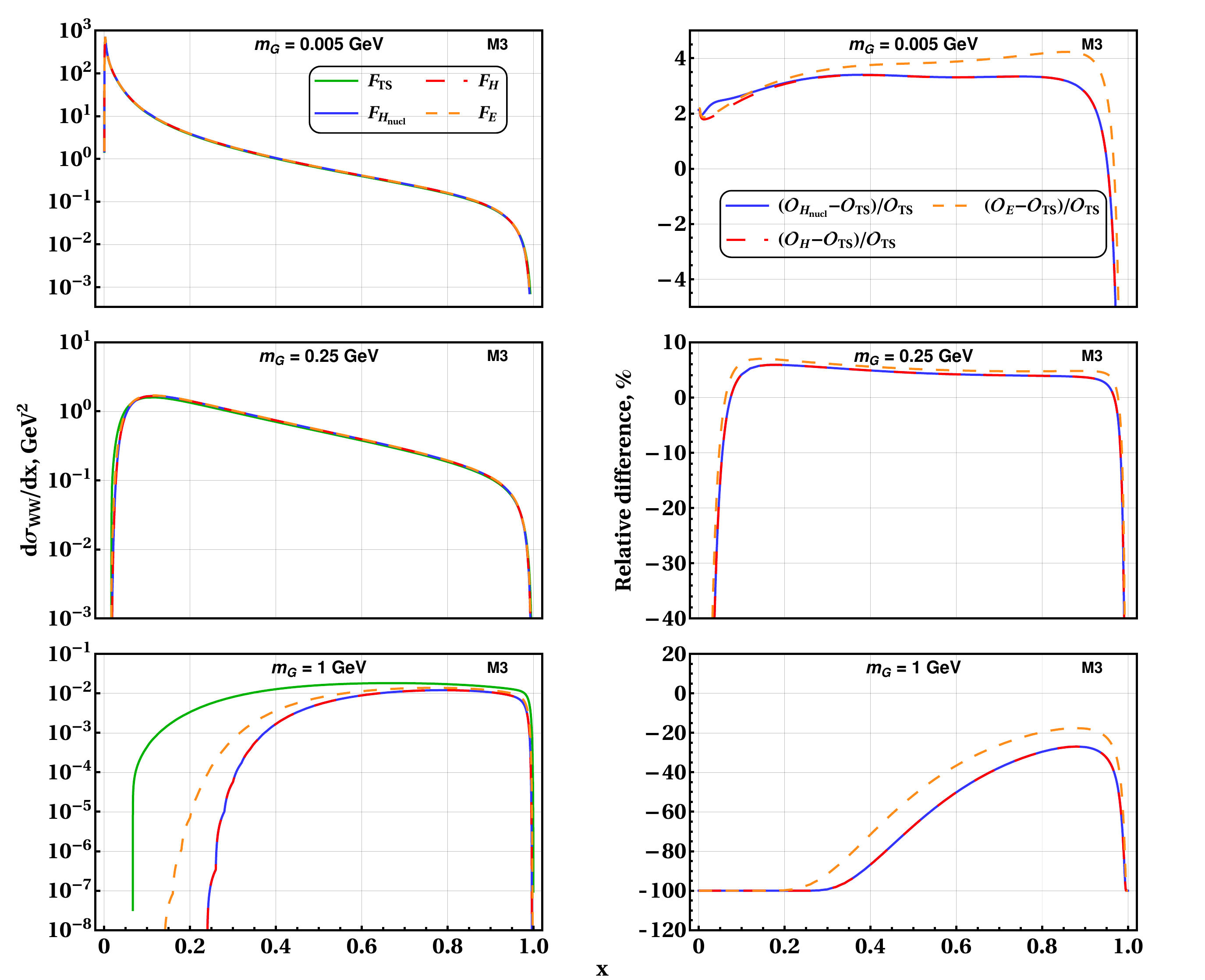}
    \caption{Same as in Fig.~\ref{fig:dsdx_NA64e} but for the M$^3$ experiment.
    \label{fig:dsdx_M3} }
\end{figure*}

\begin{figure*}[!tbh]
    \centering
    \includegraphics[width=0.65\textwidth]{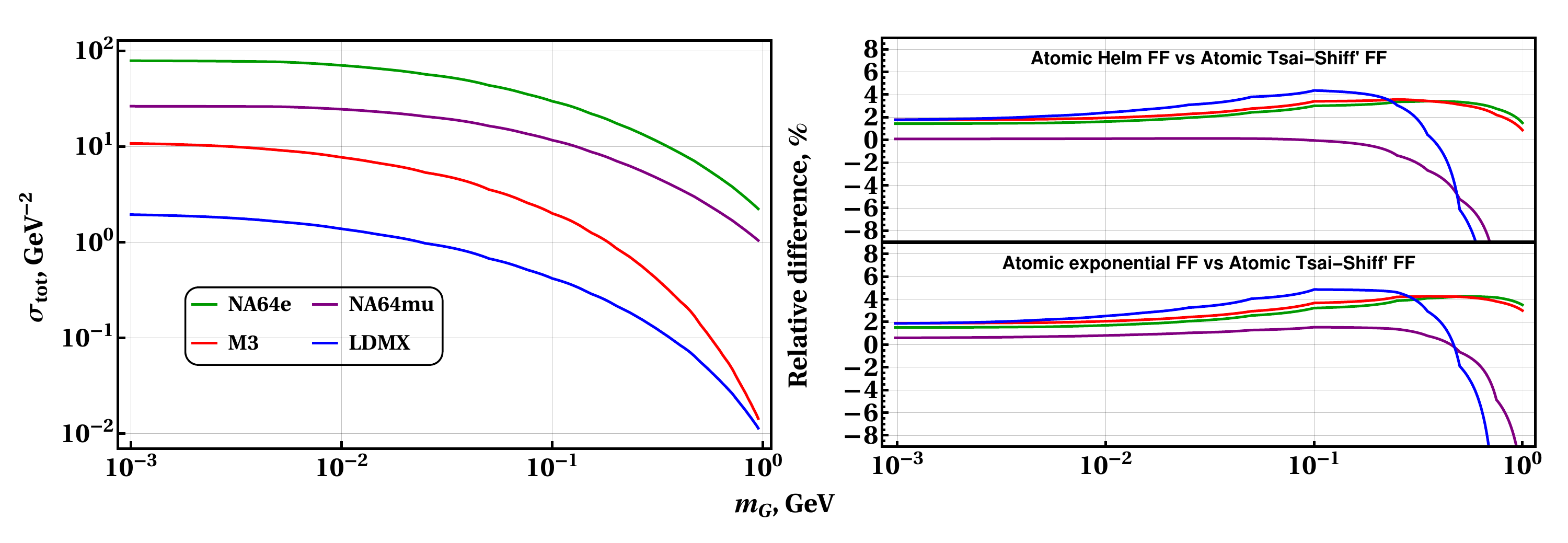}
    \caption{Left panel: total cross section of $G$-boson production as function of  its mass 
    for the set of experiments associated with benchmark Tsai-Shiff's form-factor. 
    We set here $c_{l}/\Lambda = 1 \, \mbox{GeV}^{-1}$ and integrate the cross-section over  specific 
    experimental cut range, $x_{min} \lesssim x\lesssim x_{max}$ (see text for detail).
    Right panel: the relative difference is expressed as  $(\mathcal{O}_{H} - \mathcal{O}_{TS}) / \mathcal{O}_{TS}$ (right upper) and
    $(\mathcal{O}_{E} - \mathcal{O}_{TS}) / \mathcal{O}_{TS}$ (right bottom) for atomic Helm's and the exponential 
    form-factors respectively, where~$\mathcal{O}=\sigma_{tot}$. 
    The green line corresponds to the  NA64$e$ facility, the violet line is associated with the NA64$\mu$ experiment,
    the red line corresponds to the M$^3$ facility, the blue line is associated with the LDMX experiment. 
    \label{fig:ds_All} }
\end{figure*}

The NA64$\mu$ experiment is the fixed-target facility at the CERN SPS that searches for the dark sector 
particles in the muon beam missing momentum mode $\mu N \to \mu N G$.  It can be considered as a 
complementary experiment to  NA64$e$. For  the   sensitivity estimation of NA64$\mu$ we set the energy of the muon beam 
 to be  $E_\mu\simeq 160\, \mbox{GeV}$ and  choose  $\mbox{MOT}\simeq 5\times 10^{13}$ as the projected statistics for the
 muons  accumulated on target.    The NA64$\mu$ experiment  employs the lead Shashlyk-type electromagnetic calorimeter
 that  serves as a target with a   thickness of $L_T\simeq 40 X_0 \simeq 22.5\, \mbox{cm}$. We note that one can neglect
 the  muon stopping loss in the   lead target of   $\simeq 22.5\, \mbox{cm}$  for the ultra-relativistic muons of 
 $E_\mu \simeq 160\, \mbox{GeV}$   due to small energy attenuation in the lead 
 $\langle d E_\mu /dx  \rangle \simeq 12.7\times 10^{-3} \, \mbox{GeV}/\mbox{cm}$ (see e.~g.~Ref.~\cite{Chen:2017awl} for detail).  

 In order to measure the momentum of incident and outgoing muon, the NA64$\mu$ facility employs 
 two magnet spectrometers. We choose the typical cut for the outgoing muon as 
 $E_{\mu}^{rec} \lesssim 0.5 E_\mu\simeq 80\, \mbox{GeV}$, that corresponds to $x_{min}=0.5$ in Eq.~(\ref{eq:NradGrav}).
 We note that for NA64$\mu$ facility approximately $\simeq 120$ 
 days are needed to accumulate statistics of $\mbox{MOT} \simeq 5\times 10^{13}$  relative to 
 $\mbox{EOT}\simeq 5\times 10^{12}$ for NA64$e$ facility.  This can be explained by  the increased
intensity of muon beam line at M3 that is higher by factor of $\simeq 10$ than the intensity of electron beam at H4. 

\subsection{M$^3$}

The muon missing momentum experiment (M$^3$) at Fermilab is the projected fixed target facility that aims probing dark
sector  particles by employing the muon-specific  missing energy signature $\mu N \to \mu N G(\to \chi \bar{\chi})$.
It utilizes the muon beam of $E_\mu \simeq 15\, \mbox{GeV}$ impinging on the tungsten target (W) 
($\rho=19.3\,\mbox{g cm}^{-3}, A=184\,\mbox{g mole}^{-1}, Z=74$) of the typical thickness of 
$L_{T}\simeq 50 X_0\simeq 17.5 \, \mbox{cm}$, where $X_0\simeq 0.35~\mbox{cm}$ is the radiation length of electron 
in the tungsten. The regarding facility also exploits a downstream detector to veto the Standard model background. 
It aims to collect $\mbox{MOT}\simeq 10^3$ during $\simeq 3$ months of the experimental running. The cut on the 
missing momentum of muon is chosen to be $E^{rec}_\mu \lesssim 9\, \mbox{GeV}$, this yields the lower limit on the 
energy fraction $x_{min}\simeq 0.4$ in Eq.~(\ref{eq:NradGrav}).

To conclude this subsection we note that 
for muons of $E_\mu \simeq 15\, \mbox{GeV}$ the muon stopping loss in the tungsten ~\cite{Kahn:2018cqs} is estimated to be of the order of $530\, \mbox{MeV}$ through the target medium of $50 X_0$. This allows one to neglect 
$ \langle d E_\mu /dx \rangle$ in the numerical calculation of the yield of $G$-boson at M$^3$.  As a result that 
justifies the exploiting of Eq.~(\ref{eq:NradGrav}) for the estimate of $N_G$  with $L_T\simeq 50 X_0$ 
at M$^3$ facility.

\section{The differential and total сross sections  \label{CS_WW}}

In this section we study both the single-differential and total cross sections  in the  WW 
approximation for the set of  the form-factors discussed  in Section~\ref{SectionFluxDetails}. 
The results are presented for the benchmark parameters of the lepton fixed target experiments  which can potentially 
probe the invisible  signatures associated with a lepton missing energy in bremsstrahlung-like process 
$l N \to l N G (\to \chi \overline{\chi})$.  The total  
$\sigma_{tot}$ cross-sections are  obtained by the numerical integration of the double-differential cross 
section  \eqref{dsWW} for the  specific experimental cuts.  
In the left panels of Figs.~\ref{fig:dsdx_NA64e},~\ref{fig:dsdx_LDMX},~\ref{fig:dsdx_NA64mu}, 
and~\ref{fig:dsdx_M3} we show the differential cross-sections of $G$-boson production as function of energy fraction 
$x=E_{G}/E_l$ for the NA64$e$, LDMX, NA64$\mu$ and M$^3$ experiments respectively and for the set of form-factors 
discussed in Sec.~\ref{SectionFluxDetails}.  These cross-sections are also presented for the specific set of 
$G$-boson masses, $m_G=5\, \mbox{MeV}$, $m_G=250\, \mbox{MeV}$ and $m_G=1\, \mbox{GeV}$.

Let us describe now the typical properties and the
general kinematics of $G$-boson production by the charged  lepton beams impinging on the solid target.   
It is worth mentioning that both  NA64$e$ and LDMX electron cross-sections have
a peak in the region $x \lesssim 1$ for the relatively heavy masses $m_G \gtrsim 250\, \mbox{MeV}$, it 
means that the signal of $G$-boson production is strongly forward peak, such that the dominant part of the
initial beam energy transfers to the hidden spin-2 boson. On the other hand, for the 
relatively light masses $m_G \gtrsim 5\, \mbox{MeV}$ the sharp forward peak is mitigated and the regarding 
differential cross-section is in the soft bremsstrahlung-like regime. 
It implies that the cross-section peaks in the infra-red (IR) region $x=E_G/E_e \ll 1$. 
However, for the higher  masses  $m_G \gtrsim 250\, \mbox{MeV}$  the IR peak is smeared as long as the 
energy fraction is small. 

Remarkably that both  NA64$\mu$ and M$^3$ muon cross-sections are in the soft breamsstrahlung-like regime 
$d\sigma_\mu /d x \propto \mathcal{O}(1/x)$ as long as $x\ll 1$ for all masses in the range 
$1\, \mbox{MeV}  \lesssim  m_G \lesssim 1\, \mbox{GeV}$,  since the mass of spin-2  boson is comparable to 
the muon mass in that region,  $m_G \propto  \mathcal{O}(m_\mu)$.

Let us describe now the impact of various form-factor parametrizations on the shape of the differential 
cross-sections.   In order to compare these cross-sections we choose Tsai-Shiff's 
form-factor as a benchmark one, since the latter  is exploited widely in the  calculations of WW cross-section  
for both beam dump and fixed target experiments (see
e.~g.~Refs.~\cite{Chen:2017awl,Bjorken:2009-FTE,Kahn:2018cqs,Kirpichnikov:2021jev} and references therein for detail).
In what follows in the right panels of Figs.~\ref{fig:dsdx_NA64e},~\ref{fig:dsdx_LDMX},~\ref{fig:dsdx_NA64mu}, 
and~\ref{fig:dsdx_M3} we show the relative differences between $(d \sigma/dx)_{TS}$ and the set of $(d\sigma/dx)_{H_{nucl}}$, $(d\sigma/dx)_{H}$ and $(d\sigma/dx)_{E}$ respectively for the specific 
experiments and the specific masses of $G$-boson. We recall again for clarity,
that the subscriptions of $F_{TS}(t)$, $F_{H}(t)$ and $F_E(t)$  are  related to the atomic elastic form-factors 
with screening  term,  while the label of $F_{H_{nucl}}(t)$  is associated with the nuclear elastic  
form-factor that doesn't take into account the screening effect. 

It is worth mentioning that given the mass $m_G$ all the atomic form-factor cross-sections  
for the NA64$e$ experiment  match with a reasonable accuracy at the level of $2\%-10\%$ as soon as 
$0.2 \lesssim x\lesssim 1$. However, the sizable deviation (i.~e.~at the level of $\propto \mathcal{O}(2)$)
between the nuclear and atomic cross-sections appears only for the small mass region long as $m_G \lesssim  5\, \mbox{MeV}$. 
 These effects are shown in the right panel of Fig.~\ref{fig:dsdx_NA64e}. 
 The regarding impact of the form-factor parametrization on the differential cross-section shape
 is shown in the right panel of Fig.~\ref{fig:dsdx_LDMX} for the LDMX electron fixed-target experiment. 
In addition we note that the shapes of the cross-sections can be varied for $x\ll 1$ and $m_G\lesssim 1\, \mbox{GeV}$, 
however this  energy fraction region  doesn't provide a sizeable contribution to the total cross-section of heavy masses. 

For the muon beam cross-sections of the NA64$\mu$ experiment the impact of the different form-factor parametrization 
can be estimated  at the level of $\lesssim 2\% -8\%$ for energy fraction in the range $0.2 \lesssim x\lesssim 1$.
The latter is shown in the right panel of Fig.~\ref{fig:dsdx_NA64mu}. 
 One can see also from Fig.~\ref{fig:dsdx_M3} that the shape of the differential cross-section for the M$^3$ experiment 
can be varied significantly due to the form-factors as long as $x \ll 1$ and $m_G \lesssim 1\, \mbox{GeV}$.

 Let us study now the specific shape of the  total cross-sections of $G$-boson production.
In the left panel of Fig.~\ref{fig:ds_All} we show the regarding cross-sections for the NA64$e$, LDMX, NA64$\mu$ and M$^3$
facilities, which are calculated for the benchmark Tsai-Shiff's 
atomic elastic form-factor. One can see from Fig.~\ref{fig:ds_All} 
that both NA64$e$ and NA64$\mu$ cross-sections are calculated for the lead (Pb) target ($Z\simeq 82$), nevertheless
the the total electron beam cross-section $\sigma^{tot}_e$ is generally larger by factor of $\simeq 5$ than
the muon beam cross-section $\sigma^{tot}_\mu$, even though they have a comparable energies of the impinging beams
$E_{e} \simeq E_\mu \simeq \mathcal{O} (100)\, \mbox{GeV}$. That implies the advantage of using the electron beam 
instead of muon beam at  the CERN SPS facility in the mass range 
$1\, \mbox{MeV} \lesssim  m_G \lesssim 1\, \mbox{GeV}$. 
However, if one compares the LDMX electron cross-section for the aluminium  (Al)  target ($Z\simeq 13$)
with the M$^3$ cross-section for the muon beam  impinging on the tungsten ($W$) target ($Z\simeq 74$), one can 
conclude that the regarding  advantage of using the electron beam   is  compensated by the nucleus charge 
suppression $\simeq 13/74$, even though both LDMX and M$^3$ experiments have a comparable beam energies,  
$E_\mu \simeq E_e \simeq 15 \, \mbox{GeV}$. In addition we note from the NA64$\mu$ and M$^3$ cross-sections
that the greater energy of the muon beam implies greater rate of the $G$-boson production, 
$\sigma^{tot}_{\mu}(160\,\mbox{GeV}) \gtrsim \sigma^{tot}_{\mu}(16\,\mbox{GeV})$. 
To conclude this section, we specify briefly the impact of the form-factor 
parametrization on the total cross-section. In the right panel of Fig.~\ref{fig:ds_All} the relative
differences of the total  cross-sections are shown for the set of atomic form-factors. In the small mass region 
$m_G \lesssim 100\, \mbox{MeV}$ these  differences are relatively small, i.~e.~at the level of $\lesssim 2\%$. 
However, the  regarding discrepancies can be as large as $\gtrsim 8\%$ for
the heavy mass region~$m_G \lesssim 1\, \mbox{GeV}$.

\section{The experimental limits
\label{SectionExpectedReach}}

\begin{figure*}[!tbh]
\centering
\includegraphics[width=0.65\textwidth]{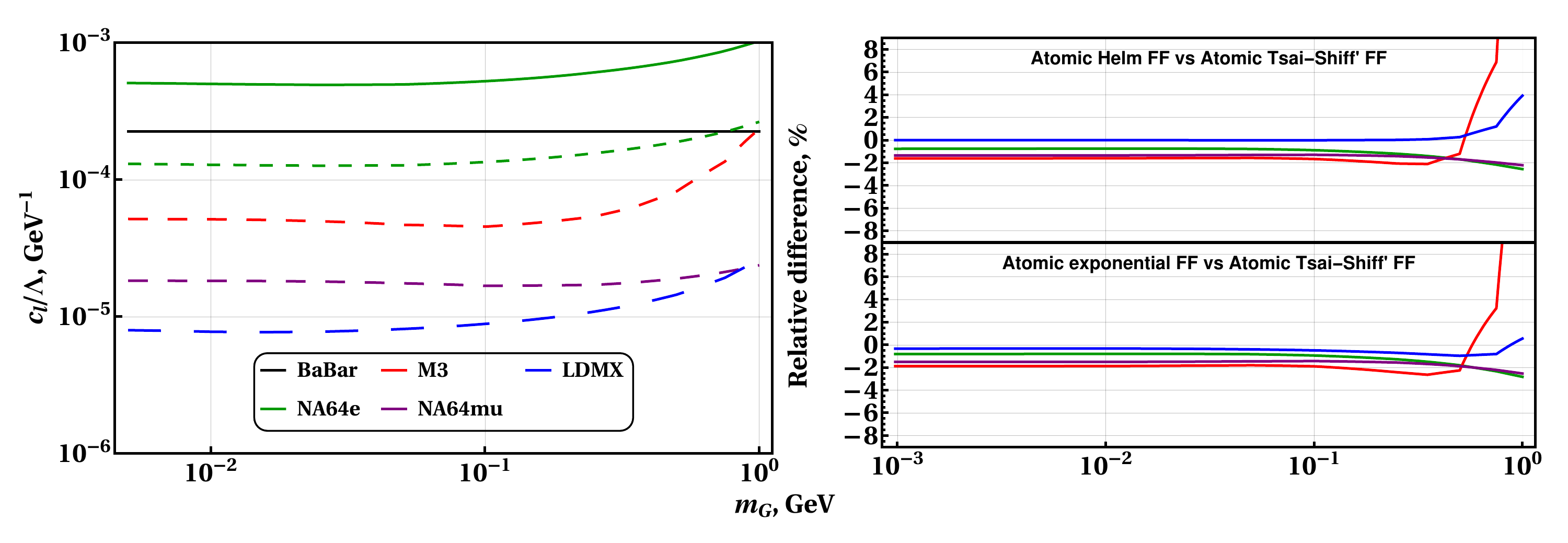}
\caption{Left panel: $90\, \%\, C.L.$ limits on $c_l/\Lambda$ coupling constant for the fixed-target and
BaBar experiment~\cite{Kang:2020-LGMDM} as a function of 
$G$-boson mass $m_G$. For all expected reaches, we imply  that 
$\mbox{Br}(G\to \chi \bar{\chi})\simeq 1$ and  
$m_G \gtrsim 2m_\chi$. In addition, in the left panel all curves for the
fixed-target facilities imply the Tsai-Shiff's form-factor in the cross-section. 
The green dashed line is the projected sensitivity of the NA64$e$ experiment for $\mbox{EOT}\simeq 5\times 10^{12}$, 
the red dashed line is the expected reach of the M$^3$ facility for $\mbox{MOT}\simeq 10^{13}$, 
the violet dashed line is the projected sensitivity of the NA64$\mu$ experiment for 
$\mbox{MOT}\simeq 5\times 10^{13}$ and the blue dashed line is the expected reach of the LDMX facility 
for $\mbox{EOT} \simeq 10^{16}$. The green solid line represents the excluded at $90\, \% \, C.L.$ bound 
of the NA64$e$ experiment for $\mbox{EOT} \simeq  3.22 \times 10^{11}$. 
Right panel: the relative difference is expressed as  $(\mathcal{O}_{H} - \mathcal{O}_{TS}) / \mathcal{O}_{TS}$ 
(right upper) and $(\mathcal{O}_{E} - \mathcal{O}_{TS}) / \mathcal{O}_{TS}$ (right bottom) for atomic Helm's and the exponential form-factors respectively, where~$\mathcal{O}=c_l /\Lambda$. 
\label{fig:EP_Range_FF} }
\end{figure*}

In this section we study current and expected reach of the lepton fixed-target experiments.  
The limit on the coupling $c_l/\Lambda$ can be obtained as follows. By exploiting both  Eq.~(\ref{eq:NradGrav})
for the number of produced $G$-bosons and the result on its production cross-section with specific 
experimental cuts, one can require that the  number of signal events is $N_G\gtrsim 2.3$.  As a result,
this yields  $90\% C.L.$ exclusion limit on the coupling constant $c_l/\Lambda$ for the background free case and null 
result of the fixed-target facilities. In the left panel of Fig.~\ref{fig:EP_Range_FF} 
we show by the dashed lines the projected sensitivities of NA64$e$, LDMX, NA64$\mu$ and M$^3$ fixed target facilities.

It is worth mentioning that the expected reach of the LDMX is rather strong, i.~e.~at the level of 
$c_l /\Lambda \gtrsim 10^{-5}\, \mbox{GeV}^{-1}$, even though the typical cross-section 
of the $G$-boson production is fairly small (see e.~g.~the blue line in the left panel of Fig.~\ref{fig:ds_All}).   
The enhanced expected reach of the LDMX is associated with fairly large number of the electrons that will be 
accumulated on target. In particular, by the final phase of data taking LDMX plans to accumulate 
$\mbox{EOT}\simeq 10^{16}$, as we discussed it above in Sec.~\ref{LDMXsubSect}.  

In addition, from the projected 
sensitivities of both NA64$\mu$ and M$^3$ muon experiments shown in Fig.~\ref{fig:EP_Range_FF}, 
one can conclude that compared to the M$^3$  option with $16\, \mbox{GeV}$ beam muons and $\mbox{MOT}\simeq 10^{13}$, the higher energy muons of 
NA64$\mu$  (e.~g.~$E_\mu \simeq 160\, \mbox{GeV}$) with $\mbox{MOT}\simeq 5 \times 10^{13}$
allow examining  a wider region in the parameter space of the spin-2 boson scenario. Nevertheless,
the ultimate projected  bounds of the NA64$e$ experiment for $\mbox{EOT}\simeq 5\times 10^{12}$ 
(see e.~g.~the green dashed line in the left panel of Fig.~\ref{fig:EP_Range_FF} 
at the level of $c_l/\Lambda \gtrsim 10^{-4}\, \mbox{GeV}^{-1}$), can be ruled out even
by the M$^3$ facility at the level of $c_l/\Lambda \gtrsim 7\times 10^{-5}\, \mbox{GeV}^{-1}$ (see e.~g.~the red 
dashed line in the left panel of Fig.~\ref{fig:EP_Range_FF}). In the left panel of Fig.~\ref{fig:EP_Range_FF}, we 
show by the solid green line the excluded limit of the NA64$e$ experiment 
(at the level of $c_l/ \Lambda \gtrsim 7\times 10^{-4}\, \mbox{GeV}^{-1}$) for the current accumulated statistics of 
$\mbox{EOT}\simeq 3.22\times 10^{11}$.  In particular,  from the analysis of the NA64$e$ data collected during  
2016-2021 runs no signal events were found for the background free case (see e.~g.~Ref.~\cite{Andreev:2022hxz}),
as we discussed it in  Sec.~\ref{NA64eSetupSect}.

It is worth mentioning that authors of Ref.~\cite{Kang:2020-LGMDM} have been already considered the bounds
 on the light spin-2 mediator from $e^+e^- \to \gamma+E_{miss}$ at BaBar experiment~\cite{BaBar:2017tiz}.
 The analysis was carried out for mainly invisible decay mode $\mbox{Br}(G\to \chi \bar{\chi})\simeq 1$. The regarding
 exclusion limit yields  the magnitude of the coupling constant
at the level of $c_e/ \Lambda \gtrsim 2\times 10^{-4}\, \mbox{GeV}^{-1}$ for the masses in the range 
   $1\,\mbox{MeV} \lesssim m_G \lesssim 1\, \mbox{GeV}$.
 These bounds are shown by the solid black line in the left panel of Fig.~\ref{fig:EP_Range_FF}. One can conclude that
 it rules out the current experimental constraints of the NA64$e$ facility for $\mbox{EOT} \simeq 3.22\times 10^{11}$.
 
To conclude this section, we discuss briefly the impact of the form-factor 
parametrization on the experimental limits $c_l/\Lambda$ for the fixed target facilities. 
In the right panel of Fig.~\ref{fig:EP_Range_FF} 
the relative differences of the coupling constants are shown for the set 
of atomic form-factors. In the small mass region  $m_G \lesssim 100\, \mbox{MeV}$ these  differences are relatively 
small, i.~e.~at the level of $\lesssim 2\%$.  However, the  regarding discrepancies can be as large as $\gtrsim 8\%$ 
for the heavy mass region $m_G \lesssim  1\, \mbox{GeV}$. 
\section{Conclusion}
In the present paper we have  discussed in detail the probing of the massive spin-2 boson $G$ trough its 
invisible decay into pair of DM particles $\chi$ with $\mbox{Br}(G\to \chi \bar{\chi})\simeq 1$. The
regarding  scenario implies that the $G$-boson serves as a mediator between charged leptons of SM and DM sector. 
The benchmark simplified coupling of this model  involves the dimension-5 operators of massive $G_{\mu \nu}$ 
field and  the  energy momentum tensors of both SM and DM particles. We have studied explicitly the missing 
energy signatures for the projected and existing lepton fixed-target experiments, such as NA64$e$, LDMX, 
NA64$\mu$ and M$^3$. Namely, by exploiting the WW approach, we have calculated $G$-boson production 
cross-section in the process $l N \to l N G$ followed by its invisible decay into DM particles 
$G\to \chi \bar{\chi}$ for the specific fixed-target experiments. We have calculated the expected reach  
for the regarding experiments, implying the background free case and null result for DM detection.   
Moreover, we have also discussed in detail the  impact of both nuclear and atomic form-factor parametrizations on
(i) the differential spectra of $G$ boson emission, (ii) the total cross-section of its production 
(iii) the experimental reach of the fixed-fixed target facilities for probing hidden spin-2 boson. 
It was found that the total yield of spin-2 boson production and the regarding experimental reach can be affected
by the form-factor at the level of $\simeq 8\%-10\%$ for the large mass region $m\lesssim 1\, \mbox{GeV}$. 
For the light masses of $G$-boson, $m_{G} \lesssim 100\, \mbox{MeV}$, the implication of the form-factors for the
yield and experimental reach is estimated at the level of $\lesssim 2\%$.

\begin{acknowledgments} 
We would like to thank  P.~Crivelli, S.~Demidov,  S.~Gninenko, D.~Gorbunov,  M.~Kirsanov, N.~Krasnikov, V.~Lyubovitskij, L.~Molina Bueno,  A.~Pukhov,  H.~Sieber and A.~Zhevlakov  
 for very helpful discussions and  correspondences.
The work of D.~V.~K on description of the  dark matter missing energy signatures of NA64$e$ 
and regarding exclusion limits for spin-2 DM mediator is  supported
by the  Russian Science Foundation  RSF grant 21-12-00379.

\end{acknowledgments}	

\appendix

\section{Matrix element for process \texorpdfstring{$l^{-} \gamma \rightarrow G l^{-}$}{Lg}} \label{MatElDeriv}

In this section, we collect an expressions for the Feynman diagrams \cite{lee:2014GMDM, PhysRevD.59.105006} associated with 
spin-2 boson production and the matrix element squared for process $l^{-} \gamma \rightarrow G l^{-}$.
In particular, we use the expression  for polarization sum of spin-2 mediator:
\begin{equation}
    \sum\limits_{s} e_{\mu \nu}(k, s) e_{\alpha \beta}(k, s)
=
    1/2( P_{\alpha \mu} P_{\beta \nu} + P_{\beta \mu} P_{\alpha \nu} - 2/3 P_{\mu \nu} P_{\alpha \beta}),
\end{equation}
where $ P_{\mu \nu} = g_{\mu \nu} - \frac{k_{\mu} k_{\nu}}{m_G^2}$. 
The vertex  for the outgoing $G$-boson in case of incoming $l(p_1)$ and outgoing $l(p_2)$ leptons is
\begin{widetext}
\begin{equation}
\begin{gathered}
    \includegraphics[width=0.25\textwidth,valign=c]{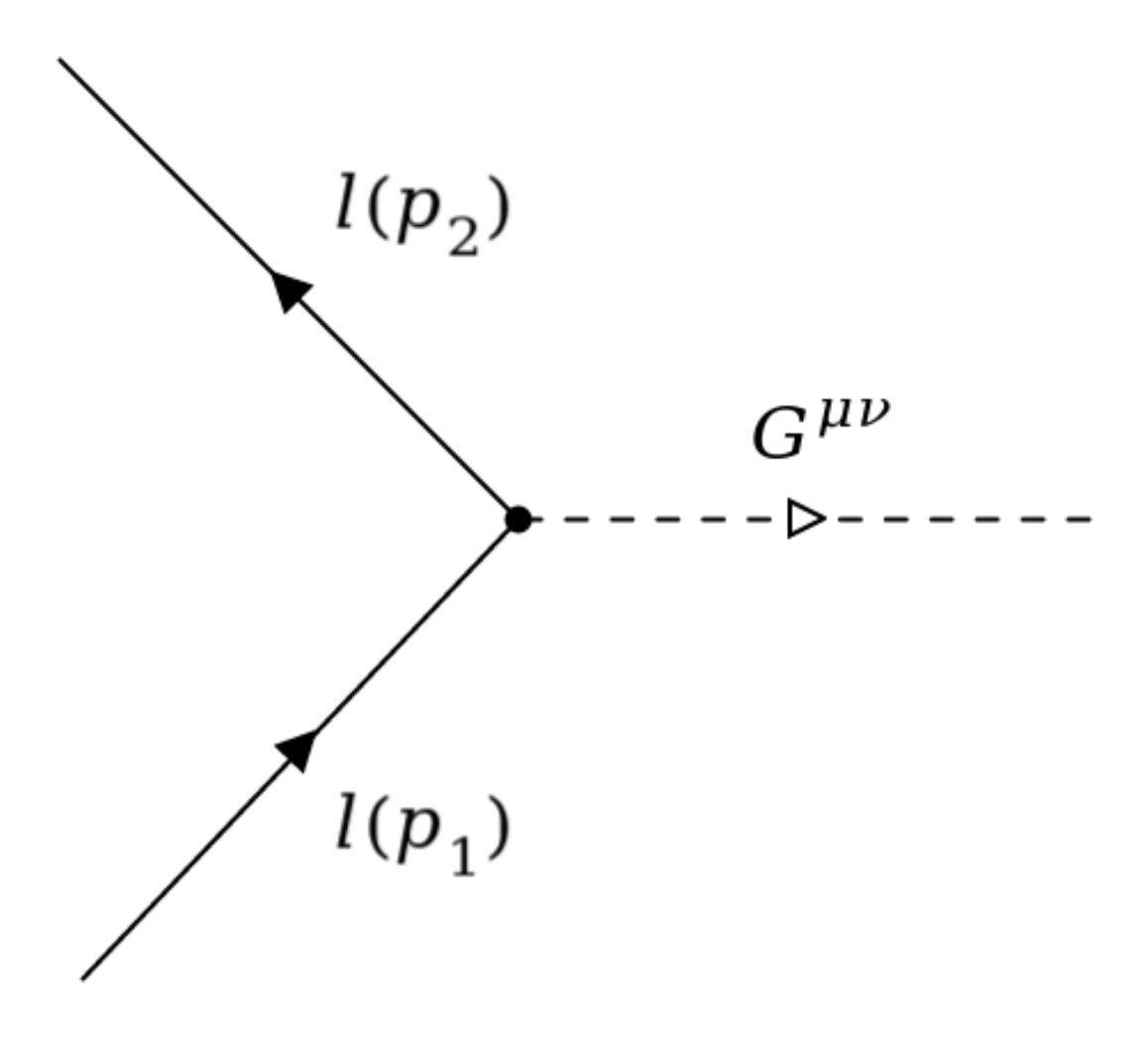} 
\end{gathered}
=
\begin{aligned}[t]
-   \frac{i c_{l}}{4 \Lambda}
    \left\lbrace    
        (p_2 + p_1)^{\mu}\gamma^{\nu} 
    +   (p_2 + p_1)^{\nu}\gamma^{\mu} 
   - \right. \\ \left. -
       2 g^{\mu \nu} ( \slashed{p_2} + \slashed{p_1} - 2 m_{l}  )
    \right\rbrace.
\end{aligned}
\end{equation}
\end{widetext}
For the incoming $V^{\alpha}(k_1)$ and outgoing $V^{\beta}(k_2)$ massless vector bosons one has
\begin{widetext}
\begin{equation}
\begin{gathered}
    \includegraphics[width=0.25\textwidth,valign=c]{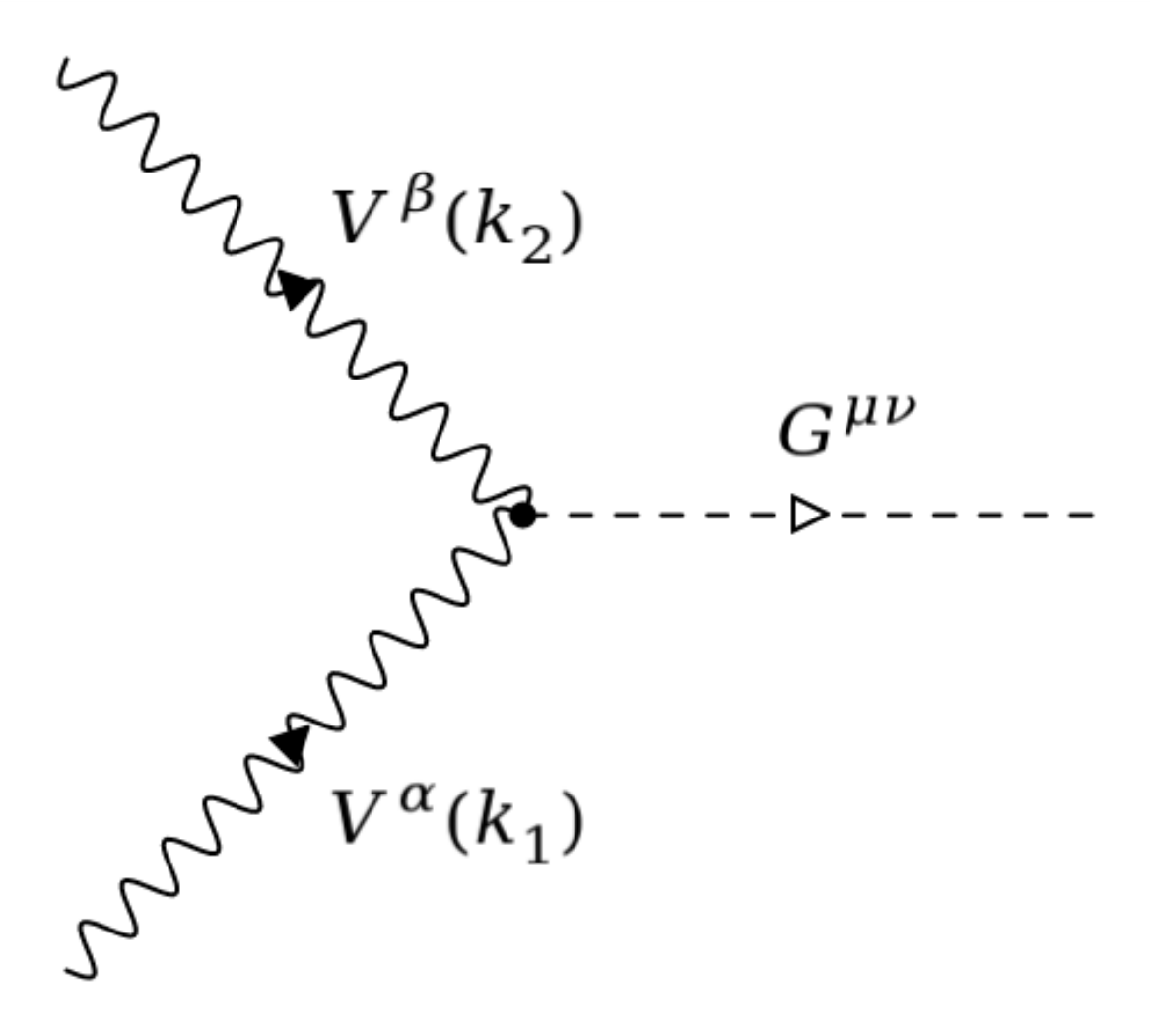} 
\end{gathered}
=
\begin{aligned}[t]
-   \frac{i c_{\gamma}}{\Lambda} 
    \left\lbrace 
        \eta^{\alpha \beta} k_1^{\mu} k_2^{\nu} + \eta^{\mu \alpha}((k_1, k_2) \eta^{\nu \beta} - k_1^{\beta} k_2^{\nu} ) 
    \right. - \\ - \left.
         \eta^{\mu \beta} k_1^{\nu} k_2^{\alpha} + 1/2 \; \eta^{\mu \nu} (k_1^{\beta} k_2^{\alpha} - (k_1, k_2) \eta^{\alpha \beta})
        + ( \mu \longleftrightarrow \nu )
    \right\rbrace.
\end{aligned}
\end{equation}
Finally the vertex for 4-point interaction is
\begin{equation}
\begin{gathered}
    \includegraphics[width=0.25\textwidth,valign=c]{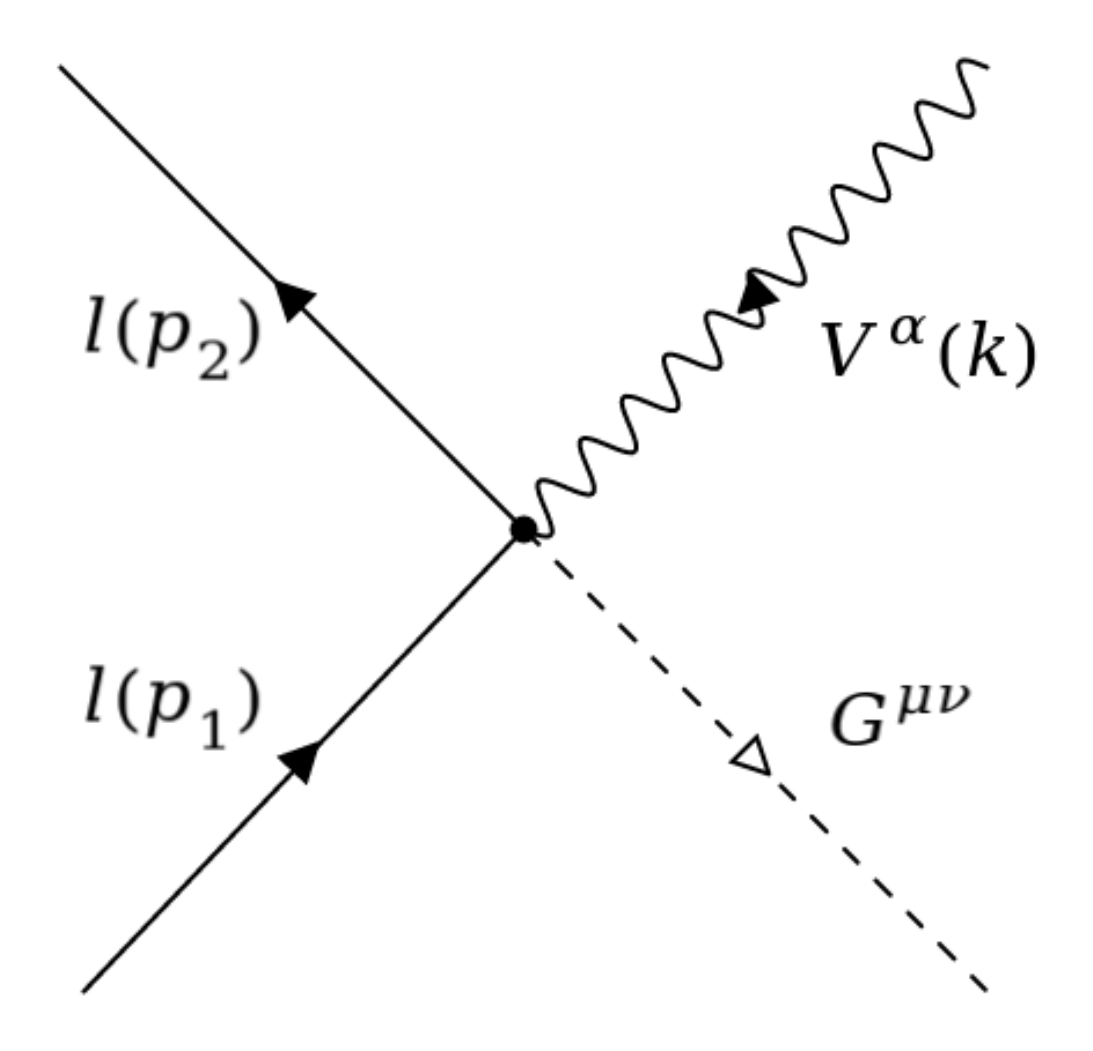} 
\end{gathered}
=
\begin{aligned}[t]
    \frac{c_{l } e}{2} 
    \left( 
        \eta^{\mu \alpha} \gamma^{\nu} + \eta^{\nu \alpha}\gamma^{\mu} 
    \right).
\end{aligned}
\end{equation}
As a result, by using FeynCalc package~\cite{Shtabovenko:2016sxi,Shtabovenko:2020gxv} 
for the Wolfram Mathematica~\cite{Mathematica}, we get matrix 
element squared for process $l^-\gamma \to l^- G$:
\begin{multline}
     \left| \mathcal{M}_{l^{-} \gamma \rightarrow G l^{-}} \right|^2 
= 
    \frac{c_l^2 e^2}{\Lambda^2}
    \frac{u_2 [s_2 - 2 m_l^2] [(t_2 + u_2)^2 + (u_2 - m_G^2)^2] [4 u_2 (2 m_l^2 - s_2) -  m_G^2 t_2] }
         {4 t_2 (u_2 - m_l^2)^2 (s_2 - m_l^2)^2}
-  \\ - 
    \frac{c_l^2 e^2}{\Lambda^2}
    \frac{m_l^2 R(m_l,m_G, t_2,u_2)}
         {12 t_2^2 (m_l^2 - u_2)^2 (s_2 - m_l^2)^2 },
\end{multline}
where  $R(m_l,m_G, t,u)$ is regular expression for $m_l$ and $m_G$:
\begin{multline}
    R(m_l, m_G, t, u)
=
        24 m_l^{10} t
    +   24 m_l^8 [m_G^4 + 5 m_G^2 t - 6 t u ] 
+ \\+   2 m_l^6 [24 m_G^6 
                + 6 m_G^4 (15 t - 8 u) 
                - 3 m_G^2 t (33 t + 92 u) 
                + 2 t (13 t^2 + 18 t u + 90 u^2)]
+ \\+  2 m_l^4 [12 m_G^8 + 12 m_G^6 (5t - 6 u) 
                + m_G^4(72 u^2 - 113 t^2 - 288 t u) 
            + \\+ 4 m_G t (t + 3 u )(13 t + 42 u) 
                - t (7t^3 + 64 t^2 u + 144 t u^2 + 240 u^3) ]
+ \\+   m_l^2 [12 m_G^8 (3 t -4 u) - 3 m_G^6 (29 t^2 + 80 t u - 48 u^2) 
                 + m_G^4 (70 t^3 + 544 t^2 u + 696 t u^2 - 96 u^3) 
            - \\ - m_G^2 t (27 t^3 + 364 t^2 u + 852 t u^2 + 912 u^3 )
                + 2 t (t^4 + 46 t^3 u + 92 t^2 u^2 + 216 t u^3  + 180 u^4)]
+ \\+   [   - 3 m_G^8 (3 t^2 + 16 t u - 8 u^2) 
            + m_G^6(15 t^3 +128 t^2 u + 168 t u^2 - 48 u^3) 
        - \\- m_G^4 (9 t^4 + 122 t^3 u + 362 t^2 u^2 +384 t u^3 - 24 u^4 )
        + \\+ m_G^2 t (3 t^4 + 68 t^3 u + 240 t^2 u^2 + 552 t u^3 + 408 u^4)
        - \\ -
            2 t u (t + u ) (7 t^3  +24 t^2 u + 72 t u^2 + 72 u^3)],
\end{multline}

\end{widetext}

\bibliography{bibl}

\end{document}